\documentclass[aps,prd,preprint,preprintnumbers,showpacs]{revtex4-1}
\usepackage{graphicx}
\usepackage{multirow}
\usepackage{amsmath}
\usepackage{slashed}
\begin{document}
\title{Decay properties of $N^*(1895)$}
\author{K.~P.~Khemchandani$^{1,4}$\footnote{kanchan.khemchandani@unifesp.br}}
\author{A.~Mart\'inez~Torres$^{2,4}$\footnote{amartine@if.usp.br}}
\author{ H.~Nagahiro$^{3, 4}$\footnote{nagahiro@rcnp.osaka-u.ac.jp} }
\author{ A.~Hosaka$^4$\footnote{hosaka@rcnp.osaka-u.ac.jp}}
\preprint{}

 \affiliation{
$^1$ Universidade Federal de S\~ao Paulo, C.P. 01302-907, S\~ao Paulo, Brazil.\\
$^2$ Universidade de Sao Paulo, Instituto de Fisica, C.P. 05389-970, Sao Paulo,  Brazil.\\
$^3$ Department of Physics, Nara Women's University,  Nara 630-8506, Japan.\\
$^4$ Research Center for Nuclear Physics (RCNP), Mihogaoka 10-1, Ibaraki 567-0047, Japan.
}

\date{\today}

\begin{abstract}
The nature of nucleon resonances is still being debated, while much experimental data are accumulated.  In this work, we focus on the negative parity resonance $N^*(1895)$ which is located in the scattering region of various meson-baryon coupled channels, and such dynamics can be crucial in understanding its properties.  To test the relevance of such hadron dynamics, we investigate the decay properties of $N^*(1895)$ in detail.  We examine how a two pole nature of $N^*(1895)$ is compatible with its observed decay properties.  Moreover, we find that the resonance decays into final states involving $\Lambda(1405)$ and $\Sigma(1400)$, where the latter is not yet observed experimentally. Such decay processes can be useful to  study the properties of the aforementioned hyperon resonances.  

\end{abstract}

\maketitle
\section{Introduction}
 The objective of the present work is to obtain the partial decay widths of  $N^*(1895)$ to light hyperon resonances, which can be useful in unraveling its nature. 
The state $N^*(1895)$ is particularly special as it is the highest mass nucleon known with $J^\pi=1/2^-$ and the particle data group (PDG)~\cite{pdg} lists all  $1/2^-$ structures found above 1800~MeV together, under the label of $N^*(1895)$. Due to this latter fact, it is unclear if one or more states correspond  to $N^*(1895)$. Indeed, in a previous work~\cite{Khemchandani:2013nma}, we found two poles with overlapping widths associated with $N^*(1895)$. The pseudoscalar/vector meson-baryon coupled channel amplitudes obtained in this former work reproduce, for example, the isospin 1/2 and  3/2 $\pi N$ amplitudes extracted from partial wave analysis~\cite{arndt} of the experimental data and the $\pi^- p \to \eta n$  and $\pi^- p \to K^0 \Lambda$  cross sections up to a total energy of about 2 GeV. 

Having the information on the poles related to $N^*(1895)$ as obtained in Ref.~\cite{Khemchandani:2013nma} using constrains from experimental data, a detailed analysis of its decay properties is important to further reveal its nature. For instance, $N^*(1895)$ cannot be described within the na\"ive quark model  \cite{Isgur:1978xj,Bijker:1994yr,Hosaka:1997kh,Takayama:1999kc}.  An $S_{11}$ resonance, within quark models based on the harmonic oscillator potential, after $N^*(1535)$ and $N^*(1650)$, is expected to appear with mass $>$ 2100 MeV \cite{Hosaka:1997kh,Takayama:1999kc}.  Hence, coupled channel hadron interactions are expected to play an important role in describing the properties of $N^*(1895)$.

In this manuscript we, thus, study the partial decay widths of $N^*(1895)$ to different pseudoscalar/vector-baryon channels and to $K \Lambda(1405)$ and $K \Sigma(1400)$ final states, where $\Lambda(1405)$ and $\Sigma(1400)$ are both $J^\pi =1/2^-$ resonances, with the former one often associated with two poles in the complex energy plane (see, for example, Ref.~\cite{osetramos,Oller:2000fj,Jido:2003cb,Hyodo:2011ur,Mai:2014xna}).  Before discussing the properties of the lesser known $\Sigma(1400)$, we would like to mention that a study of the decay processes $N^*(1895) \to K \Lambda(1405)$, $K \Sigma(1400)$ has a twofold interest: they can be useful in determining the properties of the $N^*(1895)$ as well as of $1/2^-$ light hyperons simultaneously. 
The information on the decay processes $N^*(1895)\to K \Lambda(1405)$ and  $N^*(1895)\to K \Sigma(1400)$ can also be relevant for describing the data on  $\gamma  p \to K^+ \Lambda^*,~ K^+ \Sigma^*$. In fact,  the exchange of $N^*$ resonances with masses $\geq$ 2000 MeV was found to be significant to describe the cross sections of the photoproduction of $\Lambda(1405)$ near the threshold  in Ref.~\cite{Kim:2017nxg}. Given the fact that $N^*(1895)$ lies close to the $K \Lambda(1405)$ and $K \Sigma(1400)$  thresholds, it should be important to study the contribution of  $N^*(1895)$ to the photoproduction of $\Lambda(1405)$ and $\Sigma(1400)$. The information obtained in this work can also be useful to analyze the process $\pi N \to  K^* \pi \Sigma$, which is intended to be studied at J-PARC~\cite{Noumi:2017sdz}.

Having stated the motivation of our work, we would like to dedicate a brief discussion on $\Sigma(1400)$.  There exist evidences for the existence of an isovector resonance with $J^\pi = 1/2^-$ and mass $\sim$ 1400 MeV, though with less agreement on its properties as obtained from different works \cite{Oller:2000fj,Guo,Wu:2009tu,Wu:2009nw,Gao:2010hy,Xie:2014zga,Xie:2017xwx,Khemchandani:2012ur,Roca:2013cca}. To bring a consensus on the issue, in a recent work~\cite{Khemchandani:2018amu},  we studied coupled channel  meson-baryon scattering for systems with strangeness $-1$ by determining the unknown parameters of the model using experimental data on the total cross sections of $K^- p \to K^- p$, $\bar K^0 n$, $\eta \Lambda$, $\pi^0 \Lambda$, $\pi^0 \Sigma^0$, $\pi^\pm \Sigma^\mp$ and the data on the energy level shift and width of the $1s$ state of the kaonic hydrogen. The work lead to
finding an evidence for $\Sigma(1400)$, besides $\Lambda(1405)$ and some other higher mass hyperons. In this former work, the coupled channels considered included both pseudoscalar and vector mesons. An advantage of such a treatment is that it allows us to obtain the couplings of the pseudoscalar/vector-baryon channels taken into account to the resonances found in the complex energy plane. In the present work we use the couplings determined in Ref.~\cite{Khemchandani:2018amu} to study $N^*(1895)\to K \Lambda(1405)$ and  $N^*(1895)\to K \Sigma(1400)$.

In the following section we discuss the formalism of the work where we show that the calculation of the partial widths for $N^*(1895)\to K \Lambda(1405)$ and  $N^*(1895)\to K \Sigma(1400)$ is done by considering different triangle loops involving several meson-baryon channels. In the subsequent section we present and discuss the results obtained which, we hope, are useful for experimental investigations of  $N^*(1895)$ as well as for the study of the photoproduction of $\Lambda(1405)$ and $\Sigma(1400)$.

 \section{Formalism}\label{formalism}
The  main purpose of the present work is to study the decay widths of $N^*(1895)$ to different meson-baryon channels and  final states involving unstable hyperons, in particular, $\Lambda(1405)$ and $\Sigma(1400)$. We take this opportunity to present the results on the branching ratios for $N^*(1895)$ decaying to different pseudoscalar/vector-baryon channels and compare them with the available experimental values listed by the PDG.  To study these decay processes, we rely on our previous works  on the nonstrange \cite{Khemchandani:2013nma} and on the strangeness $-1$ \cite{Khemchandani:2018amu}  meson-baryon coupled systems, where $N^*(1895)$, $\Lambda(1405)$ and $\Sigma(1400)$  appear as poles in the complex energy plane of the\ corresponding amplitudes. 

\subsection{ $N^*(1895)$, $\Lambda(1405)$ and $\Sigma(1400)$ as resonances in coupled channel dynamics }
In Ref.~\cite{Khemchandani:2013nma}, we studied the nonstrange meson-baryon  dynamics, considering the coupled channels $\pi N$, $\eta N$, $ K \Lambda$, $K \Sigma$, $\rho N$, $\omega N$, $\phi N$, $K^* \Sigma$ and  $K^* \Lambda$. The parameters of the model in this former work were fixed by making a $\chi^2$-fit to the total cross sections for $\pi^- p \to \eta n$, $K^0 \Lambda$, and the $\pi N$ scattering amplitudes, in isospin 1/2 and  3/2, known from the partial wave analysis of the related experimental data. The study lead to the finding of poles associated with $N^*(1535)$, $N^*(1650)$,  $N^*(1895)$ and $\Delta(1620)$. In this former work, two poles with overlapping widths were identified with $N^*(1895)$ (summarized in Table~\ref{poles} of the present manuscript), which interfere and, depending on the channel, produce a peak on the real axis around 1890-1910 MeV and width around 100-150 MeV. These findings are in good agreement with the values of the mass and width ($M = 1890$~to~$1930$ MeV and $\Gamma=80$~to~140~MeV, respectively) listed by the PDG~\cite{pdg}. 

The coupled channels considered in the study of meson-baryon systems with total strangeness $-1$ in Ref.~\cite{Khemchandani:2018amu} are $\pi \Sigma$, $\pi \Lambda$, $\bar K N$, $\eta \Sigma$, $\eta \Lambda$, $K\Xi$, $\rho \Sigma$, $\rho \Lambda$, $\bar K^* N$, $\omega \Sigma$, $\omega \Lambda$, $\phi \Sigma$, $\phi \Lambda$ and $K^* \Xi$. In this case too, the model parameters were constrained through $\chi^2$-fitting,  using the cross section data on the following processes: $K^- p \to K^- p$, $\bar K^0 n$, $\eta \Lambda$, $\pi^0 \Lambda$, $\pi^0 \Sigma^0$, $\pi^\pm \Sigma^\mp$. Data on the energy level shift and width of the $1s$ state of the kaonic hydrogen were also considered in Ref.~\cite{Khemchandani:2018amu}. As a result, two sets of fits of similar quality were found, denoted as ``Fit~I'' and ``Fit~II'' in Ref.~\cite{Khemchandani:2018amu}. In case of  Fit~I, two close lying poles appeared around 1400 MeV in the isovector amplitudes, while in Fit~II one pole was found with isospin $1$ around 1400 MeV.  The state related to these poles was represented as $\Sigma(1400)$.  Thus, both fits implied the presence of $\Sigma(1400)$, one indicating a possible double pole nature of the state while  the other relating a single pole to it.  However, only one of the two poles of Fit~I was found to be stable under changes in the lowest order amplitudes used in the model, such as the consideration (or not) of the contributions originating from the u-channel interaction (see Ref.~\cite{Khemchandani:2018amu} for more details). 
This latter pole is very similar to the single pole found in Fit~II. In the present work, we, thus, use the pole position found in Fit~II  of Ref.~\cite{Khemchandani:2018amu} for describing the properties of $\Sigma(1400)$. In the two sets of fits obtained in Ref.~\cite{Khemchandani:2018amu},  a double pole associated with $\Lambda(1405)$ was found, in agreement with the analysis~\cite{Roca:2013cca,Mai:2014xna,Lu:2013nza} of the data on the  electroproduction and photoproduction of $\Lambda(1405)$. Since the quality of Fit~I and II of Ref.~\cite{Khemchandani:2018amu} was similar, and, as mentioned above, we are going to use the results of Fit~II for $\Sigma(1400)$, for consistency, we use the results of the same fit for describing the properties of $\Lambda(1405)$. For convenience of the reader, the aforementioned pole positions of $\Lambda(1405)$ and $\Sigma(1400)$ are given in Table~\ref{poles} of the present manuscript.
\begin{table}[h]
\caption{The poles related to $N^*(1895)$, $\Lambda(1405)$ and $\Sigma(1400)$ as obtained in Refs.~\cite{Khemchandani:2013nma,Khemchandani:2018amu}. Notice that two poles are associated with $N^*(1895)$ and $\Lambda(1405)$.}\label{poles}
%\begin{ruledtabular}
\begin{tabular}{ccc}
\hline\hline
State&  \multicolumn{2}{c}{Pole position (MeV)}\\
& \multicolumn{2}{c}{ $E- i \Gamma/2$}\\\hline
$N^*(1895)$ & $1801-i96\quad$&$1912-i54$\\
$\Lambda(1405)$&$1385-i124\quad$&$1426-i15$\\
$\Sigma(1400)$&\multicolumn{2}{c}{$1399 - i 36$}\\\hline\hline
\end{tabular}
%\end{ruledtabular}
\end{table}

The findings of Refs.~\cite{Khemchandani:2013nma,Khemchandani:2018amu} allowed us to consider that  the transition amplitudes among the different meson-baryon channels in the vicinity of a pole can be expressed in terms of a scattering matrix $T_{ij}$ as 
\begin{equation}
T_{ij} = \frac{g_i g_j}{z - z_R},
\end{equation}
where $z_R$ corresponds to the pole position associated with the resonance in the complex plane and $g_i g_j$ is the product of the couplings of the resonance to channels $i$ and $j$, and  can be determined by calculating the residue of $T_{ij}$. In Refs.~\cite{Khemchandani:2013nma,Khemchandani:2018amu}, we obtained the couplings of $N^*(1895)$, $\Lambda(1405)$ and $\Sigma(1400)$ to the different related coupled channels. Using these couplings, the partial decay widths of $N^*(1895)$ to different pseudoscalar/baryon channels can be calculated in a straightforward way. The calculation of the amplitudes, and, consequently, the decay widths, for the processes $N^{*+}(1895) \to K^+ \Lambda(1405)$ and $N^{*+}(1895) \to K^+ \Sigma^0(1400)$ is more complex, as we discuss in the following section.

\subsection{Decay amplitudes of $N^*(1895)\to K \Lambda(1405), ~ K \Sigma(1400)$}
\begin{figure}[h!]
\includegraphics[width=0.7\textwidth]{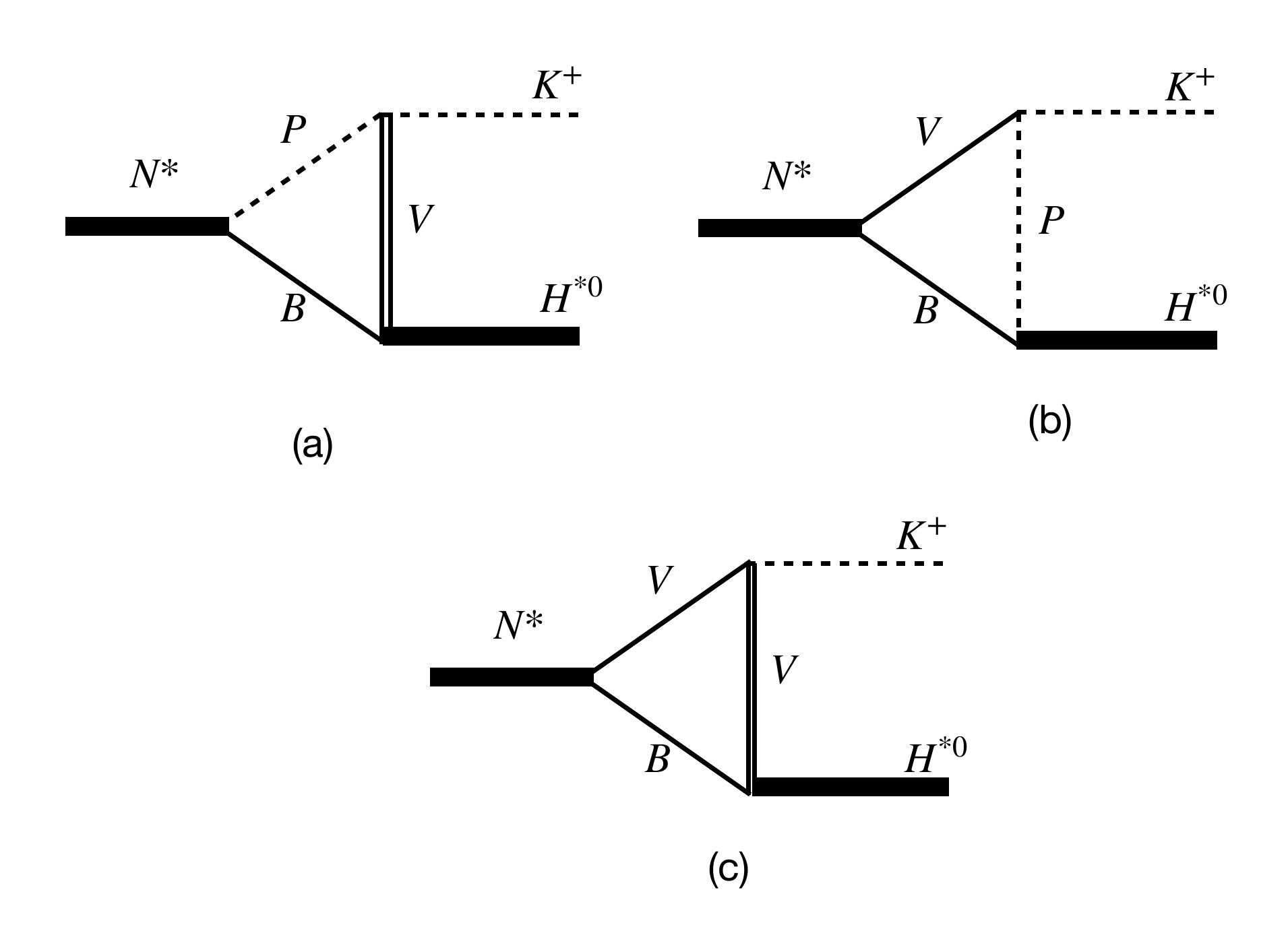}
\caption{Diagrams contributing to $N^{*+} \to K^+ H^*$, where $H^*$ refers to $\Lambda(1405)$ or $\Sigma^0(1400)$.}\label{nstar}
\end{figure}

 Based on the properties found in Refs.~\cite{Khemchandani:2013nma,Khemchandani:2018amu} for  $N^*(1895)$, $\Lambda(1405)$ and $\Sigma(1400)$, the decay processes $N^*(1895)\to K \Lambda(1405), ~ K \Sigma(1400)$ proceed through the diagrams shown in Fig.~\ref{nstar}. 
 
To obtain the amplitudes for the diagrams in Fig.~\ref{nstar}, we use the following  Lagrangians for the vertices involving mesons~\cite{Bando:1984ej,Bando:1987br}:
\begin{align}
&\mathcal{L}_{PPV}=-i g_{PPV}\langle V^\mu \left[ P, \partial_\mu P \right] \rangle, \label{vpp} \\
&\mathcal{L}_{VVP}=\frac{g_{VVP}}{\sqrt{2}}\epsilon^{\mu \nu \alpha \beta} \langle \partial_\mu V^\nu \partial_\alpha V_\beta P \rangle, \label{vvp}
\end{align}
where the couplings are related to the pion decay constant and the vector meson mass as
\begin{align}
&g_{PPV}=\frac{m_v}{2 f_\pi},\nonumber\\
&g_{VVP}=\frac{3m_v^2}{16 \pi^2 f_\pi^3},\nonumber
\end{align}
and the matrices for the mesons are
\begin{align} \nonumber
P =
\left( \begin{array}{ccc}
\dfrac{\pi^0}{\sqrt{2}} + \dfrac{\eta}{\sqrt{6}} & \pi^+ & K^{+}\\\vspace{0.2cm}
%&& \\
\pi^-& -\dfrac{\pi^0}{\sqrt{2}} + \dfrac{\eta}{\sqrt{6}} & K^{0}\\
%&&\\\vspace{0.2cm}
K^{-} &\bar{K}^{0} & \dfrac{-2\eta }{\sqrt{6}} 
\end{array}\right),~~
V^\mu =
\left( \begin{array}{ccc}
\frac{\rho^0 + \omega}{\sqrt{2}} & \rho^+ & K^{*^+}\\\vspace{0.2cm}
%&& \\
\rho^-& \frac{-\rho^0 + \omega}{\sqrt{2}} & K^{*^0}\\\vspace{0.2cm}
%&&\\
K^{*^-} &\bar{K}^{*^0} & \phi 
\end{array}\right)^\mu.
\end{align}
For the vertices involving baryons, we set effective Lagrangians which are compatible with the conventions followed in Refs.~\cite{Khemchandani:2013nma,Khemchandani:2018amu} such that we can use the couplings of the resonances to meson-baryon channels obtained in these former works,
\begin{align}
&\mathcal{L}_{N^*PB}=i g_{PBN^*} \bar B N^* P^\dagger , \nonumber\\
&\mathcal{L}_{N^*VB}=-i\frac{g_{VBN^*}}{\sqrt{3}} \bar B \gamma_5 \gamma_\mu N^* V^{\mu^\dagger} , \nonumber\\
&\mathcal{L}_{PBH^*}=g_{PBH^*} P \bar H^* B , \nonumber\\
&\mathcal{L}_{VBH^*}=i\frac{g_{VBH^*}}{\sqrt{3}} V^{\mu} \bar H^*  \gamma_\mu \gamma_5 B .\label{lageff}
\end{align}
The field $H^*$ in Eqs.~(\ref{lageff}) represents $\Sigma(1400)$ or $\Lambda(1405)$, and the couplings $g_{PBN^*}$, $g_{VBN^*}$,   $g_{PBH^*}$, $g_{VBH^*}$ are taken from Refs.~\cite{Khemchandani:2013nma,Khemchandani:2018amu}. The factor $\sqrt{3}$ in the Lagrangians for the vertices involving a vector meson is due to the fact that the Breit-Wigner amplitudes in Refs.~\cite{Khemchandani:2013nma,Khemchandani:2018amu}, for  spin $1/2$ of the $VB$ system, are written in terms of the $g_{VBB^*}$ couplings  as  
\begin{align}
 T_{{\rm VB} \to B^*\to {\rm V'B'}}\equiv \left(i g_{{\rm VBB^*}}\right)\frac{1}{\sqrt{s}-M_{B^*}+i\Gamma_{B^*}/2}\left(-ig_{\rm V'B'B^*}\right).\label{VBparam}
 \end{align}
Note that Eq.~(\ref{lageff}) leads to a spin dependent VB $\to$ VB amplitude 
\begin{align}
 T_{{\rm VB} \to B^*\to {\rm V'B'}}= \frac{1}{3}\frac{g_{{\rm VBB^*}}^2}{\sqrt{s}-M_{B^*}+i\Gamma_{B^*}/2} ~\vec \sigma \cdot \vec\epsilon_2~\vec \sigma\cdot\vec \epsilon_1, \label{vbvbBW}
\end{align}
such that, when projected on spin 1/2, it becomes
\begin{align}
 T^{s=1/2}_{{\rm VB} \to B^*\to {\rm V'B'}}= \frac{g_{{\rm VBB^*}}^2}{\sqrt{s}-M_{B^*}+i\Gamma_{B^*}/2}, \label{vbvbBW}
\end{align}
in agreement with Eq.~(\ref{VBparam}).
 
\begin{figure}[h!]
\includegraphics[width=0.4\textwidth]{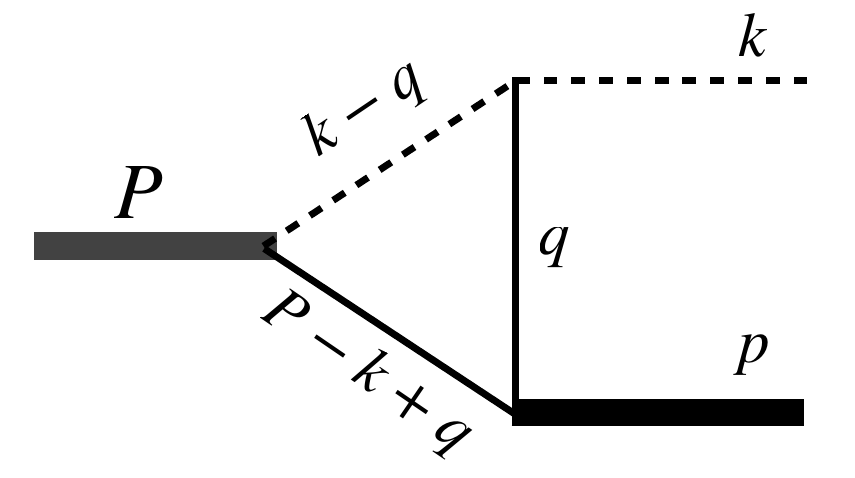}
\caption{Four-momentum labels for the particles involved in the $N^* \to K^+ \Sigma^*$ process. }\label{momentum}
\end{figure}

Having discussed the Lagrangians for the different vertices necessary to describe the decay of $N^*(1895)$ to $K^+ H^{*0}$, we can now start calculating the amplitudes for the different diagrams shown in Fig.~\ref{nstar}. We begin by writing the amplitude for the diagram in Fig.~\ref{nstar}(a) 
\begin{align}
t_a = &i  \sum_j g_{VBH^*\!, j}~ g_{PBN^*\!, j} ~g_{PPV} ~C_j ~\bar u_{H^*}\left(p\right)\gamma_\nu \gamma_5 \int\frac{d^4q}{(2\pi)^4}\Biggl\{ \frac{\left(\slashed{P}-\slashed{k}+\slashed{q}+m_{Bj}\right)}{\left(P-k+q\right)^2-m^2_{Bj}+i\epsilon}\Biggr.\nonumber\\
&\times\left.\frac{\left(-g^{\nu\mu}+\dfrac{q^\nu q^\mu}{m^2_{Vj}}\right)}{q^2-m^2_{Vj}+i\epsilon}\frac{\left(2k -q\right)_\mu}{\left(k-q\right)^2-m^2_{Pj}+i\epsilon}\right\} u_{N^*}\left(P\right),\label{ta1}
\end{align}
where we have followed the four momentum attribution  shown in Fig.~\ref{momentum}. 
 The summation over the index $j$, in Eq.~(\ref{ta1}), refers to considering different  three hadron channels in the triangle loop which can contribute to the diagram in Fig.~\ref{momentum}a. The list of such three-hadrons channels  is given in Table~\ref{Table:1} in the Appendix~\ref{ApA}.  Further, the constant $C_j$ in Eq.~(\ref{ta1}) is a coefficient obtained by performing the trace in Eq.~(\ref{vpp}) for the VPP vertex  and $m_{Bj}$, $m_{Vj}$, $m_{Pj}$ are the masses of the baryon, vector and pseudoscalar meson, respectively, corresponding to the $j$th channel in the  triangular loop. The  values of the $C_j$ coefficients are also given in Table~\ref{Table:1} in the Appendix~\ref{ApA} for each three-hadron loop present in the diagram of Fig.~\ref{nstar}a.

The product of the spinors, gamma matrices and the numerator of the expression within the curly brackets in  Eq.~(\ref{ta1}) can be worked out as
\begin{align}
N_a\left(q\right)=&\left(4 ~k\cdot p -2 ~p\cdot q-q^2\right)\bar u_{H^*}\!\left(p\right)\gamma_5u_{N^*}\!\left(P\right)-2\left(M_{H^*}+m_{Bj}\right)\bar u_{H^*}\!\left(p\right)\slashed k \gamma_5u_{N^*}\!\left(P\right)\nonumber\\
&\left(M_{H^*}+m_{Bj}\right)\bar u_{H^*}\!\left(p\right)\slashed q \gamma_5u_{N^*}\!\left(P\right)+2~\bar u_{H^*}\!\left(p\right)\slashed k\slashed q \gamma_5u_{N^*}\!\left(P\right)+\left(\frac{2 ~k\cdot q-q^2}{m^2_{vj}}\right)\nonumber\\
&\times\Bigl[\left(M_{H^*}+m_{Bj}\right)\bar u_{H^*}\!\left(p\right)\slashed q \gamma_5u_{N^*}\!\left(P\right)-\left(2~p \cdot q+q^2\right)\bar u_{H^*}\!\left(p\right) \gamma_5u_{N^*}\!\left(P\right)\Bigr],\label{Na}
\end{align}
with $M_{H^*}$ denoting the mass of $H^*$.  The integration on $dq^0$ in Eq.~(\ref{ta1}) can be done analytically by using Cauchy's theorem. It is then convenient to rewrite Eq.~(\ref{Na}) showing its explicit dependence on $q^0$. By doing so Eq.~(\ref{ta1}) becomes 
\begin{align}
t_a &= i \sum_j g_{VBH^*\!,j}~g_{PBN^*\!,j} ~g_{PPV}\mathcal{N}_{H^*}\mathcal{N}_{N^*}C_j \int\frac{d^4q}{(2\pi)^4} \Biggl\{\chi^\dagger\Bigl(\sum\limits_{i=0}^4\mathcal{A}_{i,j}[q^0]^{i}\Bigr)\chi\!\Biggr\}\nonumber\\
&\quad\times\!\frac{1}{\left[\left(P-k+q\right)^2-m^2_{Bj}+i\epsilon\right]\!\left[q^2-m^2_{vj}+i\epsilon\right]\!\left[\left(k-q\right)^2-m^2_{pj}+i\epsilon\right]}\!,\label{tabis}
\end{align}
where $\chi^\dagger$, $\chi$ correspond to the two-component spinors of $H^*$ and $N^*$, respectively. 
  The factors $\mathcal{N}_{H^*}$, $\mathcal{N}_{N^*}$ in Eq.~(\ref{tabis}) are related to the normalization of the Dirac spinors for  $H^*$ and $N^*$ 
\begin{align}
&\mathcal{N}_{H^*}=\sqrt{\frac{E_{H^*}+M_{H^*}}{2M_{H^*}}},\quad \mathcal{N}_{N^*}=\sqrt{\frac{E_{N^*}+M_{N^*}}{2M_{N^*}}},
\end{align}
where, although, $\mathcal{N}_{N^*}$ is unity in the centre of mass frame we still keep it in the equations for completeness.  The definitions of $\mathcal{A}_{i,j}$'s are as given below. The subscript $i$ on $\mathcal{A}_{i,j}$ refers to the power of $q_0$ multiplied to $\mathcal{A}_{i,j}$  and the index $j$ indicates the three-hadron channel in the loop. Defining the four-momenta in the centre of mass frame as: $P=(\sqrt{s},0)$, $k=(k^0,\vec{k})$, $p=(\sqrt{s}-k^0,-\vec{k})$ and $q=(q^0,\vec{q}~)$, we can write the expressions for  $\mathcal{A}_{i,j}$ as
\begin{align}
\mathcal{A}_{0,j}&=\vec \sigma\cdot\vec{k}\Biggl\{2\left(M_{H^*}+m_{Bj}\right)+\frac{1}{E_{H^*}+M_{H^*}}\Biggl[\Biggr.2k^0\left(M_{H^*}+m_{Bj}+2E_{H^*} \right)-2\vec k\cdot\vec q+|~\vec q~ |^2+4 |~\vec k~|^2\Biggr.\nonumber\\
&+\left.\left.\frac{|\,\vec q\,|^4+4\left(\vec k\cdot\vec q\right)^2-4\left(\vec k\cdot\vec q\right)|\,\vec q\,|^2}{m^2_{vj}}\right]\right\}-\vec \sigma\cdot \vec q\left\{ \left(M_{H^*}+m_{Bj}\right)\left(1-\frac{2\vec k\cdot\vec q-|\,\vec q\,|^2}{m^2_{vj}} \right)\right.\nonumber\\
&+2k^0+\left.2\frac{|~\vec k~|^2}{E_{H^*}+M_{H^*}}\right\},
\end{align}
\begin{align}
\mathcal{A}_{1,j}&=-\vec \sigma\cdot \vec q \left\{\frac{2k^0\left(M_{H^*}+m_{Bj}\right)}{m^2_{vj}}\right\}+\vec \sigma\cdot\vec{k}\Biggl\{2+\frac{2k^0-2E_{H^*}-M_{H^*}-m_{Bj}}{E_{H^*}+M_{H^*}}\Biggr.\nonumber\\
&+\Biggl.\left(\frac{2k^0-M_{H^*}-m_{Bj}-2E_{H^*}}{E_{H^*}+M_{H^*}}\right)\left(\frac{-2\vec k\cdot\vec q+|\vec q|^2}{m^2_{vj}}\right)\Biggr\},\\
\mathcal{A}_{2,j}&=\frac{\vec \sigma\cdot\vec{k}}{E_{H^*}+M_{H^*}}\left\{-1-\frac{1}{m^2_{vj}}\left[4k^0E_{H^*}+2k^0(M_{H^*}+m_{Bj})+2\left(-2\vec k\cdot\vec q+|\vec q|^2\right)\right]\right\}\nonumber\\
&+\vec \sigma\cdot \vec q ~\frac{\left(M_{H^*}+m_{Bj}\right)}{m^2_{vj}},
\end{align}
\begin{align}
\mathcal{A}_{3,j}&=\frac{\vec \sigma\cdot\vec{k}}{E_{H^*}+M_{H^*}}\left\{\frac{-2k^0+2E_{H^*}+m_{Bj}+M_{H^*}}{m^2_{vj}}\right\}
\end{align}
and
\begin{align}
\mathcal{A}_{4,j}&=\frac{\vec \sigma\cdot\vec{k}}{\left(E_{H^*}+M_{H^*}\right)m^2_{vj}}~.
\end{align}

The integration on the $q^0$ variable  can be done analytically, to obtain an expression like
\begin{align}
t_a =& i \sum_j g_{VBH^*\!,j}~ g_{PBN^*\!,j} ~g_{PPV}~C_j \mathcal{N}_{H^*}\mathcal{N}_{N^*}\int d\Omega_q \int\limits_{0}^{\Lambda} \frac{d|\,\vec q\,|}{(2\pi)^3} |\,\vec q\,|^2 \sum_{i=0}^{4}\chi^\dagger\Bigl[\mathcal{A}_{i,j}(\,\vec {q}\,)\Bigr]\chi\nonumber\\
&\times \left(\frac{-i N_{i,j}(\,\vec {q}\,)}{\mathcal{D}_j(\,\vec {q}\,)}\right),\label{ta2}
\end{align}
with 
\begin{align}
\frac{-iN_{i,j}(\,\vec {q}\,)}{\mathcal{D}_j(\,\vec {q}\,)}\equiv\int\frac{dq^0}{\left(2\pi\right)} \frac{\left(q^0\right)^i}{\left[\left(P-k+q\right)^2-m^2_{Bj}+i\epsilon\right]\left[q^2-m^2_{vj}+i\epsilon\right]\left[\left(k-q\right)^2-m^2_{pj}+i\epsilon\right]}.\label{nbyd}
\end{align}
A cut-off $\Lambda\simeq 600-700$ MeV is used in the integration on the three-momentum to be consistent with the work in Refs.~\cite{Khemchandani:2018amu,Khemchandani:2013nma}. The variation  of $\Lambda$ in this range allows us to estimate the uncertainties of our results. The analytical expressions for $N_{i,j}$ and $D_j$ are given in the Appendix~\ref{ApB}. 

To proceed further, we recall that the decay $N^* \to K^+ H^*$ occurs in $p$-wave and we, thus, need to write the final state projected on the partial wave $l$=1. Following Ref.~\cite{Oller:2018zts}, we write a state of two particles with spins $S_1$, $S_2$, with the centre of mass momentum $\vec k$, projected on a partial wave $l$ as
\begin{align}
\mid k, l S, J \mu\rangle =&\frac{1}{\sqrt{4\pi}} \int d\hat k \sum_{m_1, m_2} C\left(S_1, S_2, S\mid m_1, m_2, M\right) C\left(l, S, J\mid \mu-M, M, \mu\right) Y_{l\left(\mu-M\right)}\left(\hat k\right)\nonumber \\
&\times\mid \vec k, S_1 S_2, m_1 m_2 \rangle,\label{pw}
\end{align}
where $S, J$ and $M, \mu$ represent the total spin, total angular momentum and their $z$-components, respectively. Using Eq.~(\ref{pw}) and denoting the spins of $H^*$ and $N^*$ as $S_{H^*}$ and $S_{N^*}$ and their third components as $m_{H^*}$ and $m_{N^*}$, we can write the amplitude for diagram in Fig.~\ref{nstar}a, for $m_{N^*}=1/2$ (the amplitude for $m_{N^*}=-1/2$ can be obtained analogously) as
\begin{align}
\langle k,l=1, S_{\Sigma^*}, S_{N^*}\mid t_a \mid S_{N^*}, m_{N^*}=1/2 \rangle &=\frac{1}{2}\int\limits_{-1}^1 d\mathrm{cos}\theta~ \Biggl\{-\mathrm{cos} \theta~ \langle \vec k, m_{H^*}=1/2 \mid t_a\mid m_{N^*}=1/2 \rangle \nonumber\\
&-\mathrm{sin}\theta ~\langle \vec k, m_{H^*}=-1/2 \mid t_a\mid m_{N^*}=1/2 \rangle \Biggr\},\label{tapw1}
\end{align}
which, from Eq.~(\ref{ta2}), can be explicitly written as
\begin{align}
\langle~\mid t_a\mid~\rangle&=i\! \sum_j\! g_{VBH^*\!,j}~ g_{PBN^*\!,j} ~g_{PPV}\mathcal{N}_{H^*}\mathcal{N}_{N^*}~C_j\Biggl\{\frac{1}{2}\int\limits_{-1}^1 d\mathrm{cos}\theta~(-\mathrm{cos}\theta) \int\! d\Omega_q \int\limits_{0}^{\Lambda}\!\frac{d|\,\vec q\,|}{(2\pi)^3} |\,\vec q\,|^2\nonumber\\
& \times\sum_{i=0}^{4}\chi^\dagger_\uparrow\Bigl[\mathcal{A}_{i,j}(~\vec {q}~)\Bigr]\chi_\uparrow\left(\frac{-i N_{i,j}(~\vec {q}~)}{\mathcal{D}_j(~\vec {q}~)}\right)+\frac{1}{2}\int\limits_{-1}^1 d\mathrm{cos}\theta~(-\mathrm{sin}\theta) \int\! d\Omega_q \int\limits_{0}^{\Lambda}\!\frac{d|\,\vec q\,|}{(2\pi)^3} |\,\vec q\,|^2\nonumber\\
& \times \sum_{i=0}^{4}\chi^\dagger_\downarrow\Bigl[\mathcal{A}_{i,j}(~\vec {q}~)\Bigr]\chi_\uparrow\left(\frac{-i N_{i,j}(~\vec {q}~)}{\mathcal{D}_j(~\vec {q}~)}\right)\Biggr\}.\label{tapw2}
\end{align}
Note that the dependence on the spin projections of $H^*$ and $N^*$ appearing in Eq.~(\ref{tapw1}) is shown as subscripts for the spinors $\chi^\dagger$ and $\chi$ in Eq.~(\ref{tapw2}). 

We can now write the amplitudes for the diagram in Fig.~\ref{nstar}b
\begin{align}
t_b=&- \sum_j g_{PBH^*\!, j}~ g_{VBN^*\!, j} ~g_{PPV} ~D_j ~\bar u_{H^*}\left(p\right) \int\frac{d^4q}{(2\pi)^4}\Biggl\{ \frac{\left(\slashed{P}-\slashed{k}+\slashed{q}+m_{Bj}\right)}{\left(P-k+q\right)^2-m^2_{Bj}+i\epsilon} \gamma_5\gamma_\nu\Biggr.\nonumber\\
&\times\left.\frac{\left(-g^{\nu\mu}+\dfrac{\left(k-q\right)^\nu \left(k-q\right)^\mu}{m^2_{vj}}\right)}{\left(k-q\right)^2-m^2_{vj}+i\epsilon}\frac{\left(k +q\right)_\mu}{q^2-m^2_{pj}+i\epsilon}\right\} u_{N^*}\left(P\right),\label{tb1}
\end{align}
and for the diagram in Fig.~\ref{nstar}c
\begin{align}
&t_c=i \sum_j g_{VBH^*\!, j}~ g_{VBN^*\!, j} \frac{g_{VVP}}{\sqrt{2}} ~F_j ~\bar u_{H^*}\left(p\right)\int\frac{d^4q}{(2\pi)^4}\left\{\epsilon^{\lambda\nu\alpha\beta}\frac{\left(-g^\sigma_\beta+\dfrac{q^\sigma q_\beta}{m^2_{vj_1}}\right)}{\left(k-q\right)^2-m^2_{vj_1}+i\epsilon} \gamma_\sigma\gamma_5\right.\nonumber\\
&\times\frac{\left(\slashed{P}-\slashed{k}+\slashed{q}+m_{Bj}\right)}{\left(P-k+q\right)^2-m^2_{Bj}+i\epsilon} \gamma_5\gamma_\mu \left.\frac{\left(-g^{\mu}_\nu+\dfrac{\left(k-q\right)^\mu \left(k-q\right)_\nu}{m^2_{vj_2}}\right)}{q^2-m^2_{vj_2}+i\epsilon}\left(k-q\right)_\lambda q^\alpha\right\} u_{N^*}\left(P\right),\label{tc1}
\end{align}
where the constants $D_j $ and $F_j$ come from the trace in Eq.~(\ref{vpp}) describing the PPV vertex in each diagram and 
$m_{vj_1}$ and $m_{vj_2}$ in Eq.~(\ref{tc1}) are the masses of the vector mesons with four momentum $k-q$ and $q$, respectively (see Fig.~\ref{momentum}). The values of  $D_j $ and $F_j$ for different channels contributing to the diagrams in Figs.~\ref{nstar}a and \ref{nstar}b are given in Tables~\ref{Table:2} and \ref{Table:3}, respectively, of the Appendix~\ref{ApA}. As in the case of the amplitude $t_a$, we can write the amplitudes $t_b$ and $t_c$ as a polynomial of $q^0$ and integrate on the $q^0$ variable to be able to write 
\begin{align}
t_b=&-\sum_j g_{PB\Sigma^*\!, j} g_{VBN^*\!, j} ~g_{PPV} ~D_j \mathcal{N}_{H^*}\mathcal{N}_{N^*} \int d\Omega_q \int\limits_{0}^{\Lambda} \frac{d|\,\vec q\,|}{(2\pi)^3} |\,\vec q\,|^2 \sum_{i=0}^{4}\chi^\dagger\Bigl[\mathcal{B}_{i,j}(~\vec {q}~)\Bigr]\chi\nonumber\\
&\times \left(\frac{-i N_{i,j}(~\vec {q}~)}{\mathcal{D}_j(~\vec {q}~)}\right),\label{tb2}
\end{align}
\begin{align}
t_c=&-2\sum_j g_{VB\Sigma^*\!, j} g_{VBN^*\!, j} \frac{g_{VVP}}{\sqrt{2}} ~F_j \mathcal{N}_{H^*}\mathcal{N}_{N^*} \int d\Omega_q \int\limits_{0}^{\Lambda} \frac{d|\,\vec q\,|}{(2\pi)^3} |\,\vec q\,|^2 \sum_{i=0}^{2}\chi^\dagger\Bigl[\mathcal{C}_{i,j}(~\vec {q}~)\Bigr]\chi\nonumber\\
&\times \left(\frac{-i N_{i,j}(~\vec {q}~)}{\mathcal{D}_j(~\vec {q}~)}\right),\label{tc2}
\end{align}
where $N_{i,j}$ and $\mathcal{D}_j$ are as given in Eqs.~(\ref{n0})-(\ref{deno}) of the Appendix~\ref{ApB}. The expressions for $\mathcal{B}_{i,j}$ and $\mathcal{C}_{i,j}$ can also  be found in the Appendix~\ref{ApB}.

Having the amplitudes $t_b$ and $t_c$, we need to project them on $p$-wave, as done for $t_a$ [see Eq.~(\ref{tapw1})] and the final amplitude for the transition $N^* \to K^+ \Sigma^*$ is  the coherent sum of the amplitudes for the three diagrams in Fig.~\ref{nstar}
\begin{align}
t_{N^*\to K H^*}=&
\langle k,l=1, S_{\Sigma^*}, S_{N^*}\mid t_a \mid S_{N^*}, m_{N^*} \rangle+\langle k,l=1, S_{\Sigma^*}, S_{N^*}\mid t_b \mid S_{N^*}, m_{N^*} \rangle\nonumber\\
&+\langle k,l=1, S_{\Sigma^*}, S_{N^*}\mid t_c \mid S_{N^*}, m_{N^*} \rangle.\label{total}
\end{align}

\section{Results and discussions}
Having obtained the amplitudes for the diagrams shown in Fig.~\ref{nstar} for the processes $N^{*+} \to K^+ \Sigma^{*0}$ and
 $N^{*+} \to K^+ \Lambda^*$, we calculate the corresponding partial decay widths as
  \begin{align}
 \Gamma_{N^* \to K H^*} = \frac{1}{32 \pi^2}\frac{|~\vec p~|\left( 4 M_{H^*} M_{N^*}\right)}{M_{N^*}^2}\frac{1}{2 S_{N^*} +1}\int d\Omega \sum_{m_{N^*}, m_{H^*}} |t_{N^* \to K H^*}|^2,\label{eq:width}
 \end{align}
where $H^*$ denotes the hyperon resonance, $\Sigma^*$ or $\Lambda^*$. 

For the sake of clarity in the presentations of the results, we represent the two poles of $N^*(1895)$ found in Ref.~\cite{Khemchandani:2013nma} as  $N_1^*(1895)$ (for the lower pole at $1801- i 96$ MeV) and $N_2^*(1895)$ (for the higher pole at $1912- i 54$ MeV). Similarly, we shall refer to the lower and upper mass poles of $\Lambda(1405)$ (see Table~\ref{poles}) as $\Lambda_1(1405)$ and $\Lambda_2(1405)$, respectively. 

Before discussing the results, it is important to mention that although the central mass value of $N_1^*(1895)$ is below the $H^*$-kaon threshold(s), the decay width $N^*_1 \to K^+ H^*$ is finite, due to the width of $N_1^*(1895)$  (see Table~\ref{poles}), which can be taken into account through the convolution of the width [given by Eq.~(\ref{eq:width})] over the varying mass of $N^*$ as
\begin{eqnarray}
\Gamma_{N^* \to K H^*} = \frac{1}{N} \int\limits_{(M_{N^*}-2\Gamma_{N^*})^2}^{(M_{N^*}+2\Gamma_{N^*})^2}
d\tilde{m}^2\!\! \left( - \dfrac{1}{\pi}\right) \text{Im} \left\{\frac{1}{\tilde{m}^2 - M_{N^*}^2 +  i M_{N^*}\Gamma_{N^*}}\right\} \Gamma_{N^* \to K H^*} (\tilde{m}). \label{tconv}
\end{eqnarray}
In Eq.~(\ref{tconv}), $\Gamma_{N^* \to K H^*}(\tilde{m})$ is calculated using Eq.~(\ref{eq:width}), with the mass of $N^*$ varying in the range $\pm~ 2 \Gamma_{N^*}$, and 
\begin{eqnarray}
N = \int\limits_{(M_{N^*}-2\Gamma_{N^*})^2}^{(M_{N^*}+2\Gamma_{N^*})^2}  d\tilde{m}^2 \left( - \dfrac{1}{\pi}\right) \text{Im} \left\{\dfrac{1}{\tilde{m}^2 - M_{N^*}^2 + i M_{N^*} \Gamma_{N^*}}\right\},
\end{eqnarray}
is a normalization factor. As a result we obtain the widths which are summarised in Table.~\ref{Table1}.
\begin{table}[h!]
\caption{Partial decay widths of $N^*(1895) \to K H^*$. The subscripts $1$, $2$ on $N^*$ and on $\Lambda$ refer to the respective lower and upper mass poles (as shown in Table.~\ref{poles}). }\label{Table1}
\begin{tabular}{cc}
\hline\hline
Decay process& Partial width (MeV) \\ 
\hline
$N^{*+}_1 \to K^+ \Lambda^*_1$&$10.4\pm1.3$\\
$N^{*+}_1\to K^+ \Lambda^*_2$&$6.4\pm0.8$\\
$N^{*+}_1\to K^+ \Sigma^{*0}$&$3.8\pm0.5$\\
$N^{*+}_2\to K^+ \Lambda^*_1$&$1.9\pm0.1$\\
$N^{*+}_2\to K^+ \Lambda^*_2$&$1.1\pm0.2$\\
$N^{*+}_2\to K^+ \Sigma^{*0}$&$4.1\pm0.4$\\
\hline\hline
\end{tabular}
\end{table}
The uncertainty in the results is determined by allowing the cut-off, $\Lambda$, on the three-momentum integration to vary in the range $600-700$ MeV. We refer the reader to Eq.~(\ref{ta2}) to look for the dependence on $\Lambda$ in the formalism. We would like to add here that the $H^*$'s also have finite decay widths, which we considered analogously to the way we take into account the width of $N^*$. We find that the widths of $H^*$'s do not practically change the results in Table~\ref{Table1}.

\begin{figure}[h!]
\centering
\includegraphics[width=0.6\textwidth]{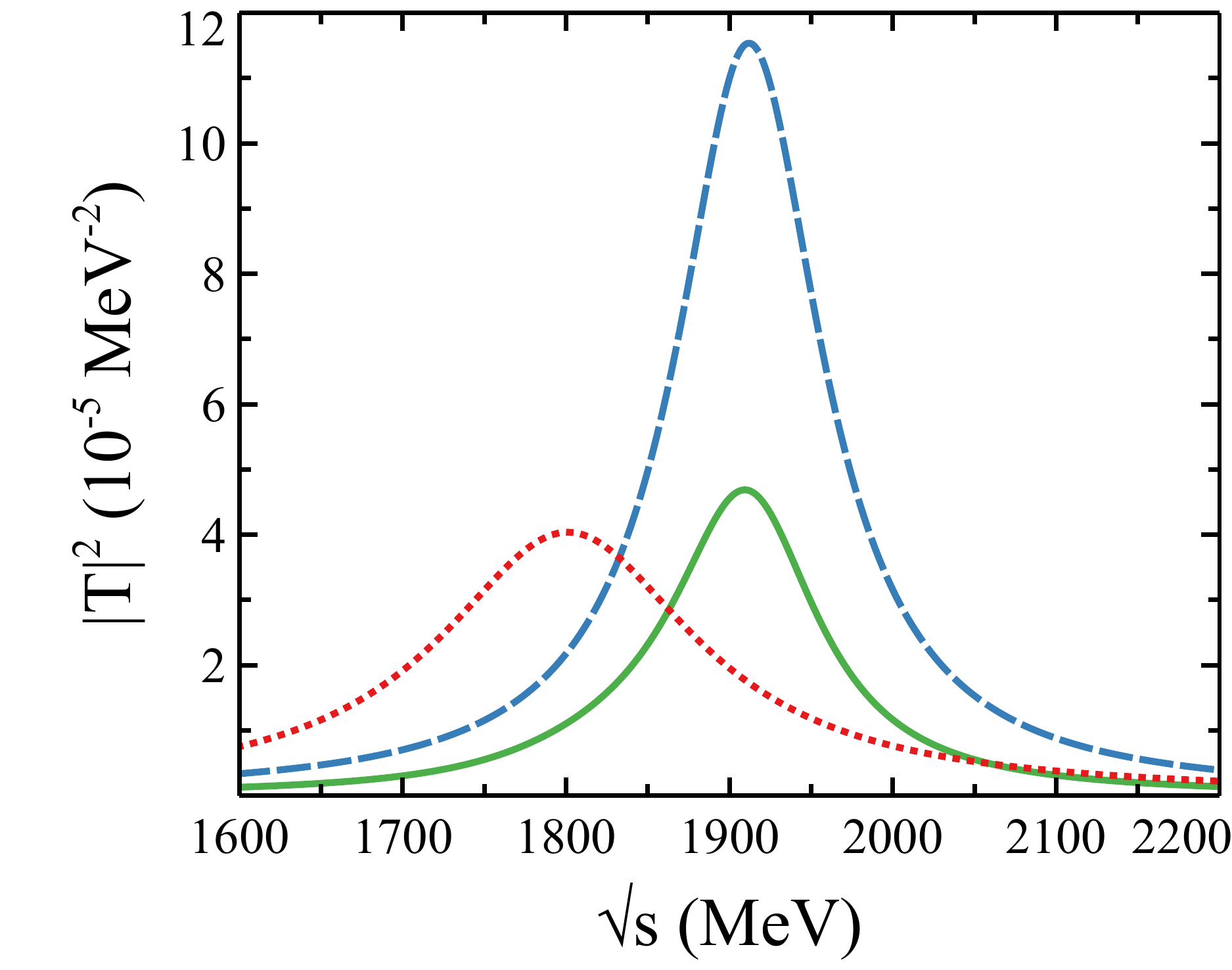}
\caption{Modulus squared amplitudes related to $N^*_1$ (dotted line), $N^*_2$ (dashed line) and their interference (solid line), which produces a unique peak, in this case, around 1900 MeV.}\label{inter}
\end{figure}

Further, it might be useful, from the experimental point of view, to provide the partial width of $N^*(1895)$ as a state on the real energy axis, produced by the superposition of the two poles in  the complex plane. To illustrate such a superposition effect, we  show the $K\Lambda\to K\Lambda$ amplitude in Fig.~\ref{inter} obtained by summing coherently the Breit-Wigners associated with the two $N^*(1895)$  poles 
\begin{align}
t_{K\Lambda}=\frac{g_{N^*_1K\Lambda}^2}{\sqrt{s}-M_{N^*_1}+i \Gamma_{N^*_1}/2}+\frac{g_{N^*_2K\Lambda}^2}{\sqrt{s}-M_{N^*_2}+i \Gamma_{N^*_2}/2},
\end{align}
where $g_{N^*_1K\Lambda}=-0.5 -i0.6$, $g_{N^*_2K\Lambda} =-0.7 +i0.3$ are taken from Ref.~\cite{Khemchandani:2013nma} and $M_{N^*_1}$, $M_{N^*_2}$, $\Gamma_{N^*_1}$, $\Gamma_{N^*_2}$ (determined in Ref.~\cite{Khemchandani:2013nma} too) are as given in Table~\ref{poles}.

To determine the decay width of  $N^*(1895)$ to $K^+\Sigma^0(1400)$, where $N^*(1895)$ is now the superposition of $N_1^*(1895)$ and $N_2^*(1895)$, we proceed in the following way: we sum the amplitudes for $N^{*+}_{1,2}\to K^+ \Sigma^0(1400)$ and use an average mass $\sim$1895 MeV and width $\sim 120$~MeV for $N^*(1895)$ in the phase space. These values correspond to the peak position and full width at the half maximum, respectively, found in the squared amplitudes on the real axis for most channels in Ref.~\cite{Khemchandani:2013nma}. As a result, we obtain
\begin{align}
\Gamma_{N^{*+}(1895)\to K^+ \Sigma^0(1400)}=\left(6.3 \pm 0.5\right)  \text{ MeV},\label{GS}\\
\text{Br}\left[{N^{*+}(1895)\to K^+ \Sigma^0(1400)}\right]=(5.3\pm0.4)\%,\label{BrS}
\end{align}
with ``Br'' representing the branching fraction.

In case of the decay to $K^+\Lambda(1405)$, we sum the amplitudes $N_1^{*+}(1895)\to K^+\Lambda_1(1405)$, $N_1^{*+}(1895)\to K^+\Lambda_2(1405)$, $N_2^{*+}(1895)\to K^+ \Lambda_1(1405)$, and $N_2^{*+}(1895) \to K^+\Lambda_2(1405)$.  A mass value of 1405 MeV is used for $\Lambda(1405)$ in the phase space.  Further, as in the calculation of the partial width of $N^*(1895) \to K^+\Sigma^0(1400)$, an average mass and width for $N^*(1895)$ have been considered in the calculation of the phase space. The values, thus, obtained are 
\begin{align}
\Gamma_{N^{*+}(1895)\to K^+ \Lambda(1405)}= \left(8.3 \pm 1.3\right) \text{ MeV},\label{GL}\\
\text{Br}\left[{N^{*+}(1895)\to K^+ \Lambda(1405)}\right]= (6.9\pm1.1)\%.\label{BrL}
\end{align}

Next, in Table~\ref{Br} 
\begin{table}[h!]
\caption{Branching ratios (in the isospin base) of the two poles of $N^*(1895)$ to different pseudoscalar-baryon and vector-baryon channels.}\label{Br}
\begin{tabular}{c|c|c|cc}
\hline
Decay channel & \multicolumn{2}{|c|}{ Branching ratios ($\%$)}  & Experimental   \\
& $N_1^*(1895)$ & $N_2^*(1895)$&data~\cite{pdg} \\\hline \hline
$\pi N$             &9.4    &  13.5     &   2-18  \\
$\eta N$           &2.7    &   22.5    &   15-40\\
$K \Lambda$   &10.9  &  24.0     &   13-23 \\
$K \Sigma$      &0.7    &   31.9    &    6-20\\
$\rho N$           &5.6    &  4.3     &     $<$18\\
$\omega N$     &25.7  &   7.6    &       16-40\\
$\phi N$             &8.9   &   2.8     &      --\\
$K^* \Lambda$ &12.1 &  25.8     &      4-9 \\
$K^* \Sigma$    &6.1   &   2.4    &      -- \\\hline
\end{tabular}
\end{table}
we provide the branching ratios for each of the two poles of $N^*(1895)$ to different PB and VB channels in the isospin base and compare them with the experimental values, whenever possible. We calculate the $N^*\to$~PB,~VB decay widths as
 \begin{align}
 \Gamma_{N^* \to \text{PB (VB)}} = \frac{|~\vec p~|}{4\pi}\frac{ M_{B}}{M_{N^*}} |g_{N^* \to \text{PB (VB)}}|^2,\label{eq:width2}
 \end{align}
and convolute over the width of $N^*$ by using Eq.~(\ref{eq:width2}) in Eq.~(\ref{tconv}). As can be seen, we obtain compatible results.  Notice that the last column of Table~\ref{Br} is a compilation of findings from the PDG~\cite{pdg}, which shows that the partial widths to the different pseudoscalar- and vector-baryon channels are of the same order in spite of the larger phase space available in the former case. Such findings from experimental data cannot be easily described within the quark model. In fact, the couplings obtained in Ref.~\cite{Khemchandani:2013nma} show that $N^*(1895)$ couples more strongly to the vector-baryon channels, which clearly indicates that the hadron dynamics plays an important role in describing the properties of $N^*(1895)$.  

 To finalize the discussions on the decay widths, it is important to consider another possible source of uncertainty present in the model which is the relative phases in the Lagrangians.  The relative phases among the Lagrangians in Eq.~(\ref{lageff}) are set as in Refs.~\cite{Khemchandani:2013nma,Khemchandani:2018amu} where the couplings of the $N^*$/$H^*$ to the PB/VB channels were determined. However, there may exist an ambiguity in the relative phase among the Lagrangians used for the meson vertices [Eqs.~(\ref{vpp}) and (\ref{vvp})]. It is then important to discuss the sensitivity of our results on the ambiguity in the relative phase of the PPV and VVP Lagrangians. In case of the $N^*(1895)$ decay to $K \Lambda(1405)$, we find that the amplitude for the diagram in Fig.~\ref{nstar}b gives the dominant contribution such that the results are basically insensitive to the relative phase among the PPV and VVP vertices. 
 For the $N^*(1895)$ decay to $K \Sigma(1400)$ the contribution of Fig.~\ref{nstar}c is such that there exists a large cancellation between the amplitudes of $N^*_1(1895)  \to K \Sigma(1400)$ and $N^*_2(1895)  \to K \Sigma(1400)$. As a consequence the decay width of the superposed $N^*(1895)$ to  $K \Sigma(1400)$ depends weakly on  the relative phase of the PPV and VVP vertices. For example, if we consider $g_{VVP} \to - g_{VVP}$ in Eq.~(\ref{vvp}) and fix the cut-off $\Lambda=700$ MeV to regularize the triangular loops, we obtain the following decay widths
  \begin{align}
\Gamma_{N^{*+}(1895)\to K^+ \Lambda(1405)}= 6.4 \text{ MeV},\\
\Gamma_{N^{*+}(1895)\to K^+ \Sigma(1400)}= 6.5 \text{ MeV},
\end{align}
which should be compared with Eqs.~(\ref{GS}) and (\ref{GL}). It can be seen that the uncertainties which can arise from such  phase ambiguities are compatible with the ones already implemented in the model.

\begin{figure}[h!]
\centering
\includegraphics[width=0.4\textwidth]{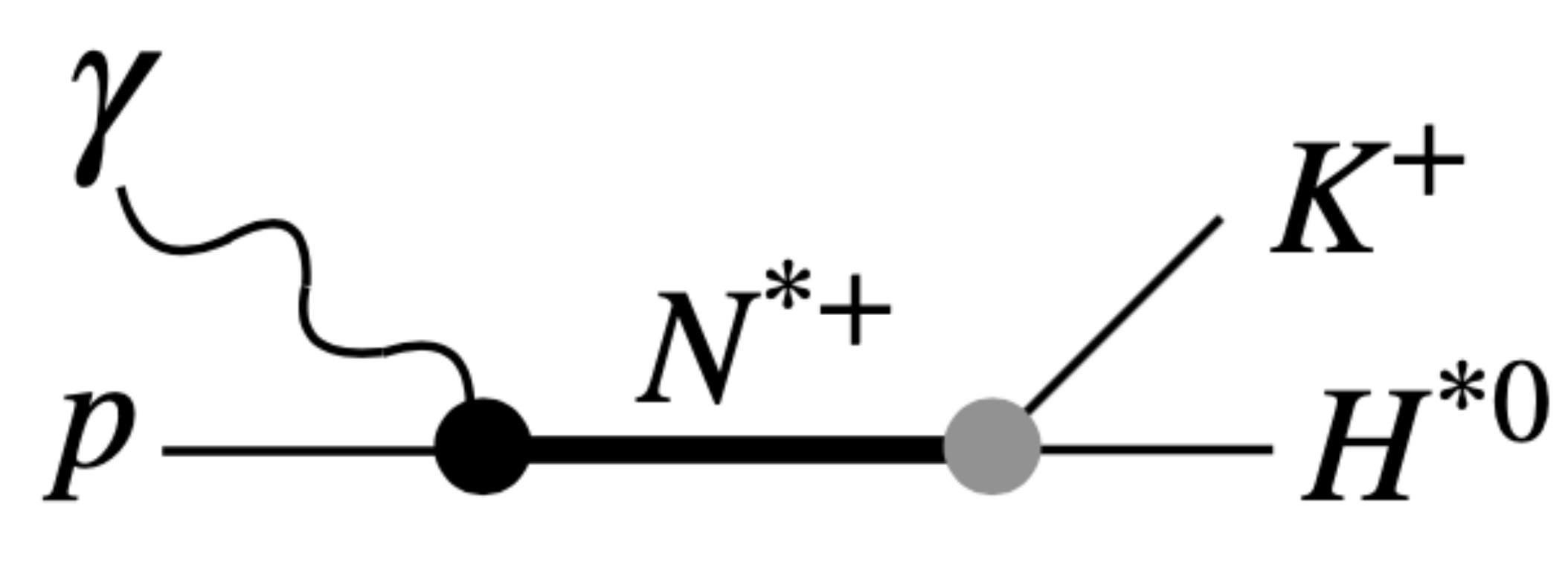}
\caption{Contribution of $N^*(1895)$ in $H^{*0}$ photoproduction, where $H^{*}$ denotes $\Lambda(1405)$ or $\Sigma(1400)$.}\label{photo}
\end{figure}

\begin{figure}[h!]
\centering
\includegraphics[width=0.45\textwidth]{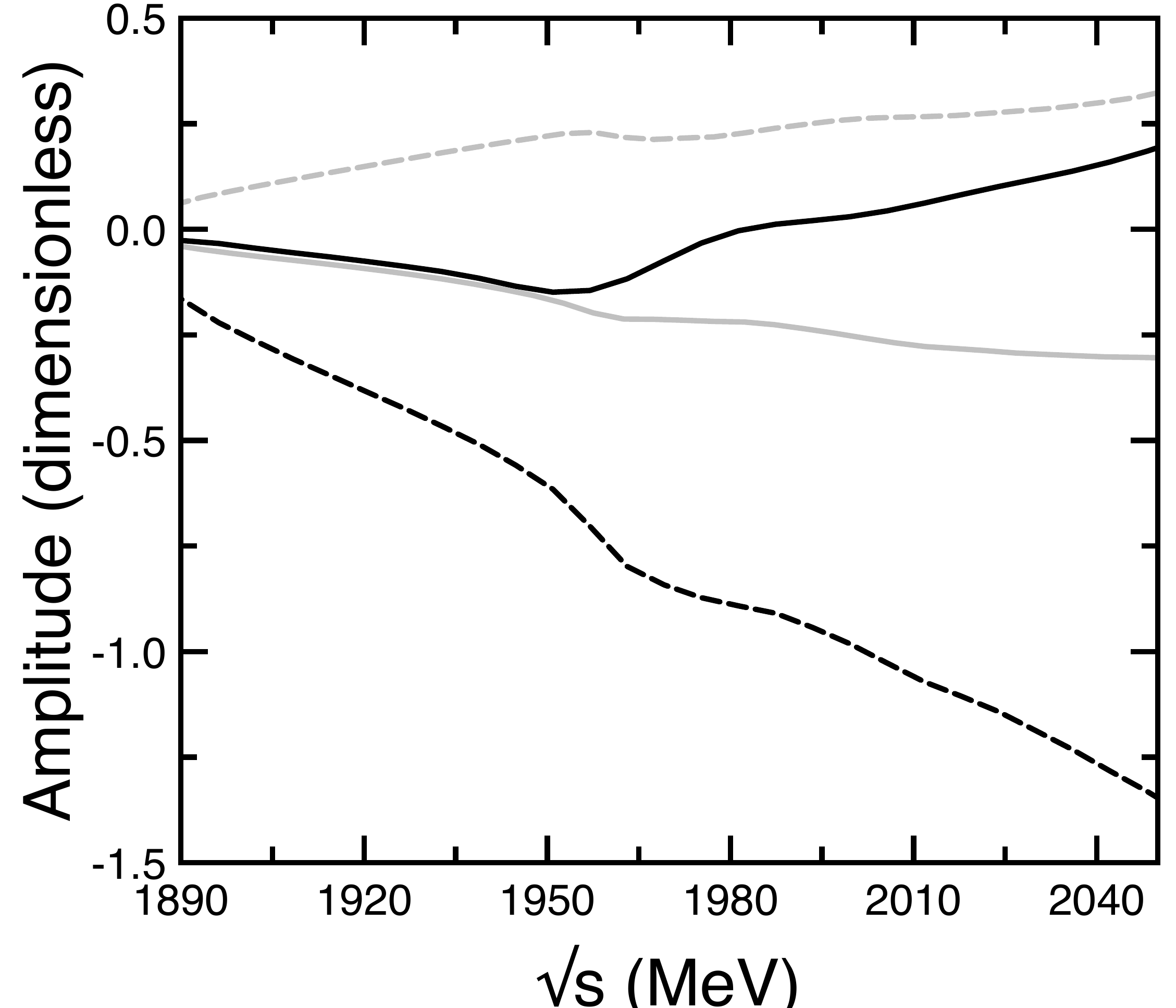}\quad\includegraphics[width=0.45\textwidth]{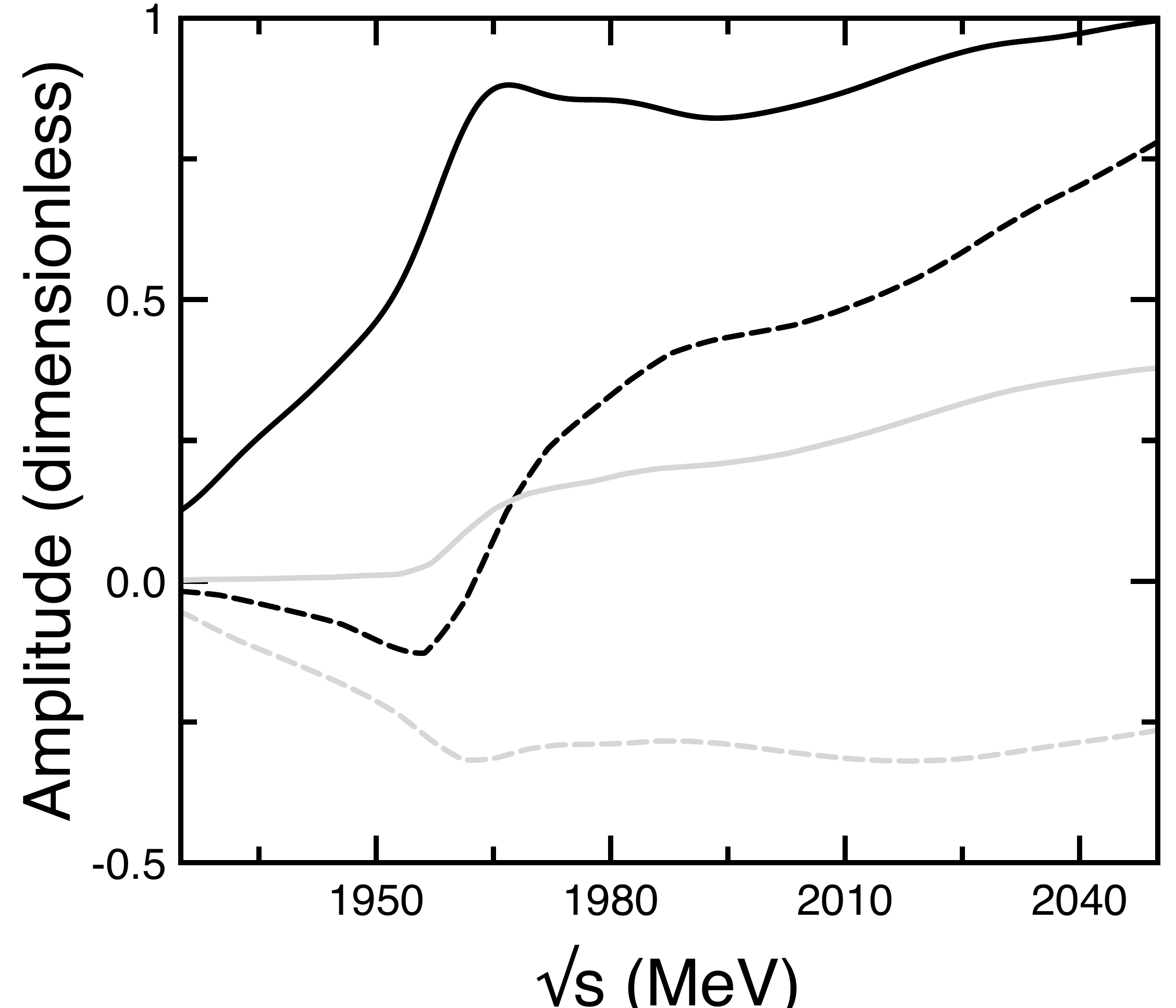}\\
\vspace{0.4cm}
\includegraphics[width=0.45\textwidth]{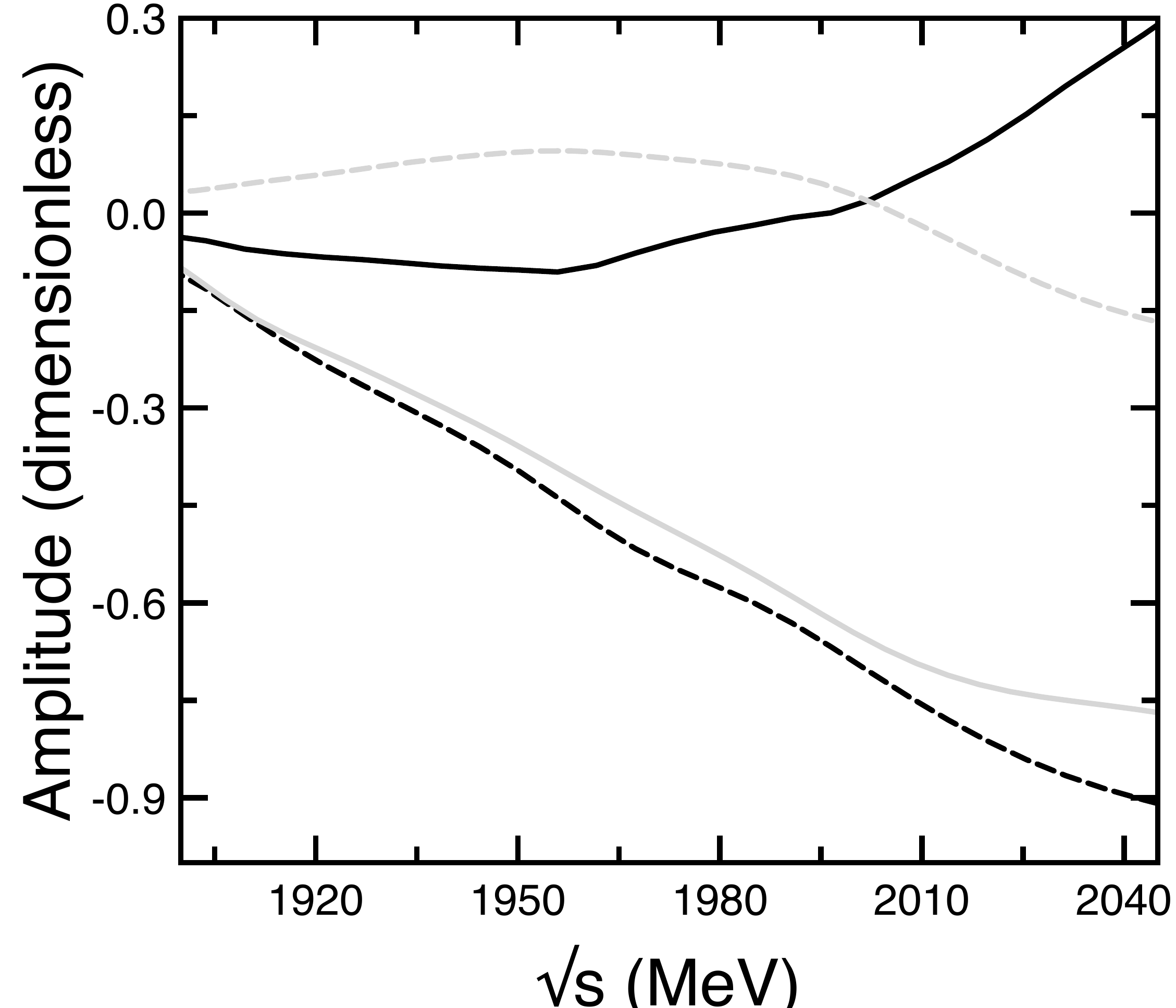}
\caption{Real (solid lines) and imaginary (dashed lines) parts of the amplitudes for the processes $N^{*+}_{1,2}\to K^+\Lambda_1$ (top panel, left side), $N^{*+}_{1,2}\to K^+\Lambda_2$ (top panel, right side) and $N^{*+}_{1,2}\to K^+\Sigma^0(1400)$ (lower panel). The dark (light) color line represents the amplitude related  to $N^*_1$ ($N^*_2$) in the initial state.}\label{tE}
\end{figure}

Finally, it can also be important to provide the energy dependence of the amplitudes obtained in this work, which  can be useful in investigating reactions where $N^*(1895)$ is produced in an intermediate state. For example,  the  process  $\gamma p\to K\Lambda(1405),~K\Sigma(1400)$ can proceed as depicted in Fig.~\ref{photo}. 
Since $N^*(1895)$ has a finite width, determining the cross sections of such a process requires the energy dependent $N^{*+}(1895)\to K^+ H^{*0}$ vertex.
Having this in mind, we show in Fig.~\ref{tE} the real (solid lines) and imaginary parts (dashed lines) of the amplitudes for the processes $N^{*+}_{1,2}\to K^+\Lambda_{1,2}$ and $K^+\Sigma^{0}(1400)$ in the energy region of interest.  
%\clearpage
\section{Summary}
  In this work we have studied the decay process of $N^*(1895)$ to channels involving light hyperon resonances, which are $K\Lambda(1405)$ and $K\Sigma(1400)$. We also provide the information on the decays of $N^*(1895)$  to various pseudoscalar- and vector-baryon channels. The formalism is based on the nature of $N^*(1895)$, $\Lambda(1405)$ and $\Sigma(1400)$ which is dominantly described in terms of meson-baryon coupled channel scattering. We find that the branching ratios obtained for decays to $K\Lambda(1405)$ and $K\Sigma(1400)$  are comparable to those for channels like $\pi N$ and $K^* \Lambda$. The branching ratios  of $N^*(1895)$ to  the channels $K\Lambda(1405)$ and $K\Sigma(1400)$  should be relevant  to describe a process, like, $\gamma  p \to K \Lambda(1405)$, on which data already exists~\cite{Moriya:2013hwg, Scheluchin:2020mhn}.  The results obtained in our work can also be useful in the analyses of other processes producing light hyperons through the exchange of  $N^*(1895)$ in the intermediate state, for example, $\pi N \to  K^* \pi \Sigma$, which is intended to be studied at JPARC~\cite{Noumi:2017sdz}. 

\section*{Acknowledgements}
K.P.K and A.M.T gratefully acknowledge the  support from the Funda\c c\~ao de Amparo \`a Pesquisa do Estado de S\~ao Paulo (FAPESP), processos n${}^\circ$ 2019/17149-3 and 2019/16924-3, by the Conselho Nacional de Desenvolvimento Cient\'ifico e Tecnol\'ogico (CNPq), grants n${}^\circ$ 305526/2019-7 and 303945/2019-2. A.M.T also thanks the partial support from mobilidade Santander for travelling to Japan (edital PRPG no 11/2019). H.N. is supported in part by Grants-in-Aid for Scientific Research (JP17K05443 (C)). AH is supported in part by Grants-in-Aid for Scientific Research (JP17K05441 (C)) and for Scientific Research on Innovative Areas (No. 18H05407).

\appendix
\section{Contributions from the isospin trace in the PPV vertices of the diagrams in Fig.~\ref{nstar} }\label{ApA}
In this appendix we provide the tables with the values of the coefficients $C_j$, $D_j$, $F_j$ appearing in the amplitudes $t_a$, $t_b$ and $t_c$, respectively, [see Eqs.~(\ref{ta1}), (\ref{tb1}) and (\ref{tc1})] corresponding  to the  different channels considered in each of the diagrams in Fig.~\ref{nstar}. We also provide the relation between the couplings given in the isospin base in Refs.~\cite{Khemchandani:2018amu,Khemchandani:2013nma} and in the charge base, which are required in the present work. To obtain these relations, we follow the phase convention: $K^- = -\mid 1/2, -1/2\rangle$, $K^{*-} = -\mid 1/2, -1/2\rangle$, $\Sigma^+ = -\mid 1, 1\rangle$, $\rho^+ = -\mid 1, 1\rangle$, and $\pi^+ = -\mid 1, 1\rangle$.
{\squeezetable\begin{table}[h!]
\caption{Different channels considered in the triangle loop in Fig.~\ref{nstar}a and the value of the $C_j$ factors in Eq.~(\ref{ta1}), together with the couplings $g_{H^* VB}$ and $g_{N^* PB}$ in the charge base in terms of those given in Refs.~\cite{Khemchandani:2018amu,Khemchandani:2013nma} in the isospin base. }\label{Table:1}
\begin{tabular}{c|c|c|c||c|c|c|c}
\hline
\multicolumn{4}{c||}{Process $N^{*+} \to K^+ \Sigma^{*0}$}&\multicolumn{4}{|c}{Process $N^{*+} \to K^+ \Lambda^*$}\\\hline
Channel in the loop &$C_{j}$& $g_{\Sigma^* VB}$ &$g_{N^* PB}$&Channel in the loop &$C_{j}$& $g_{\Lambda^* VB}$ &$g_{N^* PB}$  \\
\hline\hline
$\pi^0 p K^{*+}(892)$&$-\frac{1}{\sqrt{2}}$&$-\frac{1}{\sqrt{2}}~g_{\Sigma^* \bar K^*N}$&$-\frac{1}{\sqrt{3}}~g_{N^*\pi N}$&$\pi^0 p K^{*+}(892)$&$-\frac{1}{\sqrt{2}}$&$\frac{1}{\sqrt{2}}~g_{\Lambda^* \bar K^*N}$&$-\frac{1}{\sqrt{3}}~g_{N^*\pi N}$\\
$\pi^+ n K^{*0}(892)$&$-1$&$~\frac{1}{\sqrt{2}}~g_{\Sigma^* \bar K^*N}$&$-\sqrt{\frac{2}{3}}~g_{N^*\pi N}$&$\pi^+ n K^{*0}(892)$&$-1$&$~\frac{1}{\sqrt{2}}~g_{\Lambda^* \bar K^*N}$&$-\sqrt{\frac{2}{3}}~g_{N^*\pi N}$\\
$\eta p K^{*+}(892)$&$-\sqrt{\frac{3}{2}}$&$-\frac{1}{\sqrt{2}}~g_{\Sigma^* \bar K^*N}$&$g_{N^*\eta N}$&$\eta p K^{*+}(892)$&$-\sqrt{\frac{3}{2}}$&$\frac{1}{\sqrt{2}}~g_{\Lambda^* \bar K^*N}$&$g_{N^*\eta N}$\\
$K^+ \Lambda \rho^0$&$~\sqrt{\frac{1}{2}}$&$~g_{\Sigma^* \rho\Lambda}$&$g_{N^*K \Lambda}$&$K^+\Sigma^0\rho^0$&$\sqrt{\frac{1}{2}}$&$-\sqrt{\frac{1}{3}}~g_{\Lambda^*\rho\Sigma}$&$\sqrt{\frac{1}{3}}~g_{N^* K\Sigma}$\\
$K^0 \Sigma^+ \rho^+$&$1$&$~\frac{1}{\sqrt{2}}g_{\Sigma^* \rho\Sigma}$&$\sqrt{\frac{2}{3}}g_{N^*K \Sigma}$&$K^0 \Sigma^+ \rho^+$&1&$-\sqrt{\frac{1}{3}}~g_{\Lambda^*\rho\Sigma}$&$\sqrt{\frac{2}{3}}~g_{N^* K\Sigma}$\\
$K^+ \Sigma^0 \rho^0$&$~\frac{1}{\sqrt{2}}$&$0$&$\sqrt{\frac{1}{3}}g_{N^*K \Sigma}$&$K^+ \Lambda \omega$&$\sqrt{\frac{1}{2}}$&$g_{\Lambda^*\omega\Lambda}$&$g_{N* K \Lambda}$\\
$K^+ \Sigma^0 \omega$&$~\frac{1}{\sqrt{2}}$&$~g_{\Sigma^* \omega\Sigma}$&$\sqrt{\frac{1}{3}}g_{N^*K \Sigma}$&$K^+ \Lambda \phi$&$-1$&$g_{\Lambda^*\phi\Lambda}$&$g_{N* K \Lambda}$ \\
$K^+ \Sigma^0 \phi$&$-1$&$~g_{\Sigma^* \phi\Sigma}$&$\sqrt{\frac{1}{3}}g_{N^*K \Sigma}$&--&--&--&--\\
\hline
\end{tabular}
\end{table}
\begin{table}[h!]
\caption{Different channels considered in the triangle loop in Fig.~\ref{nstar}b and the value of the $D_j$ factors in Eq.~(\ref{tb1}), together with the couplings $g_{H^* PB}$ and $g_{N^* VB}$ in the charge base in terms of those given in Refs.~\cite{Khemchandani:2018amu,Khemchandani:2013nma} in the isospin base. }\label{Table:2}
\begin{tabular}{c|c|c|c||c|c|c|c}
\hline
\multicolumn{4}{c||}{Process $N^{*+} \to K^+ \Sigma^{*0}$}&\multicolumn{4}{|c}{Process $N^{*+} \to K^+ \Lambda^*$}\\\hline
Channel in the loop&$D_{j}$& $g_{\Sigma^* PB}$ &$g_{N^* VB}$&Channel in the loop &$D_{j}$& $g_{\Lambda^* PB}$ &$g_{N^* VB}$  \\
\hline\hline
$\rho^0 p K^{+}$&$\frac{1}{\sqrt{2}}$&$-\frac{1}{\sqrt{2}}~g_{\Sigma^* \bar KN}$&$-\frac{1}{\sqrt{3}}~g_{N^*\rho N}$&$\rho^0 p K^{+}$&$\frac{1}{\sqrt{2}}$&$\frac{1}{\sqrt{2}}~g_{\Lambda^* \bar KN}$&$-\frac{1}{\sqrt{3}}~g_{N^*\rho N}$\\
$\omega p K^{+}$&$\frac{1}{\sqrt{2}}$&$-\frac{1}{\sqrt{2}}~g_{\Sigma^* \bar KN}$&$~g_{N^*\omega N}$&$\omega p K^{+}$&$\frac{1}{\sqrt{2}}$&$\frac{1}{\sqrt{2}}~g_{\Lambda^* \bar KN}$&$~g_{N^*\omega N}$\\
$\phi p K^{+}$&$-1$&$-\frac{1}{\sqrt{2}}~g_{\Sigma^* \bar K N}$&$g_{N^*\phi N}$&$\phi p K^{+}$&$-1$&$\frac{1}{\sqrt{2}}~g_{\Lambda^* \bar K N}$&$g_{N^*\phi N}$\\
$\rho^+ n K^0$&$1$&$~\frac{1}{\sqrt{2}}~g_{\Sigma^* \bar K N}$&$-\sqrt{\frac{2}{3}}g_{N^*\rho N}$&$\rho^+ n K^0$&$1$&$~\frac{1}{\sqrt{2}}~g_{\Lambda^* \bar K N}$&$-\sqrt{\frac{2}{3}}g_{N^*\rho N}$\\
$K^{*+}(892) \Lambda \pi^0$&$-\frac{1}{\sqrt{2}}$&$g_{\Sigma^* \pi\Lambda}$&$g_{N^*K^* \Lambda}$&$K^{*+}(892) \Sigma^0 \pi^0$&$-\frac{1}{\sqrt{2}}$&$-\frac{1}{\sqrt{3}}~g_{\Lambda^*\pi \Sigma}$&$\frac{1}{\sqrt{3}}~g_{N^* K^* \Sigma}$ \\
$K^{*+}(892) \Sigma^0 \pi^0$&$-\frac{1}{\sqrt{2}}$&$0$&$\sqrt{\frac{1}{3}}g_{N^*K^* \Sigma}$&$K^{*0}(892) \Sigma^+ \pi^+$&$-1$&$-\frac{1}{\sqrt{3}}~g_{\Lambda^*\pi \Sigma}$&$\sqrt{\frac{2}{3}}~g_{N^* K^* \Sigma}$
\\
$K^{*+}(892) \Sigma^0 \eta$&$-\sqrt{\frac{3}{2}}$&$~g_{\Sigma^* \eta\Sigma}$&$\sqrt{\frac{1}{3}}g_{N^*K^* \Sigma}$&$K^{*+}(892) \Lambda^0 \eta$&$-\sqrt{\frac{3}{2}}$&$~g_{\Lambda^* \eta\Lambda}$&$~g_{N^* K^*\Lambda}$\\
$K^{*0}(892) \Sigma^+ \pi^+$&$-1$&$~\frac{1}{\sqrt{2}}g_{\Sigma^* \pi\Sigma}$&$\sqrt{\frac{2}{3}}g_{N^*K^* \Sigma}$&--&--&--&--\\
\hline
\end{tabular}
\end{table}
\begin{table}[h!]
\caption{Different channels considered in the triangle loop in Fig.~\ref{nstar}c and the value of the $F_j$ factors in Eq.~(\ref{tc1}), together with the couplings $g_{H^* VB}$  and $g_{N^* VB}$ in the charge base in terms of those given in Refs.~\cite{Khemchandani:2018amu,Khemchandani:2013nma} in the isospin base. }\label{Table:3}
\begin{tabular}{c|c|c|c||c|c|c|c}
\hline
\multicolumn{4}{c||}{Process $N^{*+} \to K^+ \Sigma^{*0}$}&\multicolumn{4}{|c}{Process $N^{*+} \to K^+ \Lambda^*$}\\\hline
Channel in the loop&$~~~F_{j}~~~$& $g_{\Sigma^* VB}$ &$g_{N^* VB}$&Channel in the loop&$~~~F_{j}~~~$& $g_{\Lambda^* VB}$ &$g_{N^* VB}$  \\
\hline\hline
$\rho^0 p K^{*+}(892)$&$\frac{1}{\sqrt{2}}$&$-\frac{1}{\sqrt{2}}~g_{\Sigma^* \bar K^*N}$&$-\frac{1}{\sqrt{3}}~g_{N^*\rho N}$&$\rho^0 p K^{*+}(892)$&$\frac{1}{\sqrt{2}}$&$\frac{1}{\sqrt{2}}~g_{\Lambda^* \bar K^*N}$&$-\frac{1}{\sqrt{3}}~g_{N^*\rho N}$\\
$\omega p K^{*+}(892)$&$\frac{1}{\sqrt{2}}$&$-\frac{1}{\sqrt{2}}~g_{\Sigma^* \bar K^*N}$&$~g_{N^*\omega N}$&$\omega p K^{*+}(892)$&$\frac{1}{\sqrt{2}}$&$\frac{1}{\sqrt{2}}~g_{\Lambda^* \bar K^*N}$&$~g_{N^*\omega N}$\\
$\phi p K^{*+}(892)$&$1$&$-\frac{1}{\sqrt{2}}~g_{\Sigma^* \bar K^*N}$&$g_{N^*\phi N}$&$\phi p K^{*+}(892)$&$1$&$\frac{1}{\sqrt{2}}~g_{\Lambda^* \bar K^*N}$&$g_{N^*\phi N}$\\
$\rho^+ n K^{*0}(892)$&$1$&$~\frac{1}{\sqrt{2}}~g_{\Sigma^* \bar K^*N}$&$-\sqrt{\frac{2}{3}}g_{N^*\rho N}$&$\rho^+ n K^{*0}(892)$&$1$&$\frac{1}{\sqrt{2}}~g_{\Lambda^* \bar K^*N}$&$-\sqrt{\frac{2}{3}}g_{N^*\rho N}$\\
$K^{*+}(892) \Lambda \rho^0$&$\frac{1}{\sqrt{2}}$&$g_{\Sigma^* \rho\Lambda}$&$g_{N^*K^* \Lambda}$&$K^{*+}(892) \Sigma^0 \rho^0$&$\frac{1}{\sqrt{2}}$&$-\frac{1}{\sqrt{3}}$$g_{\Lambda^* \rho\Sigma}$&$\frac{1}{\sqrt{3}}$$g_{N^* K^*\Sigma}$\\
$K^{*+}(892) \Sigma^0 \rho^0$&$\frac{1}{\sqrt{2}}$&$0$&$\sqrt{\frac{1}{3}}g_{N^*K^* \Sigma}$&$K^{*+}(892) \Sigma^+ \rho^+$&$1$&$-\frac{1}{\sqrt{3}}$$g_{\Lambda^* \rho\Sigma}$&$\sqrt{\frac{2}{3}}$$g_{N^* K^*\Sigma}$\\
$K^{*+}(892) \Sigma^0 \omega$&$\frac{1}{\sqrt{2}}$&$~g_{\Sigma^* \omega\Sigma}$&$\sqrt{\frac{1}{3}}g_{N^*K^* \Sigma}$&$K^{*+}(892) \Lambda \omega$&$\frac{1}{\sqrt{2}}$&$~g_{\Lambda^* \omega\Lambda}$&$g_{N^*K^* \Lambda}$\\
$K^{*+}(892) \Sigma^0 \phi$&$1$&$~g_{\Sigma^* \phi\Sigma}$&$\sqrt{\frac{1}{3}}g_{N^*K^* \Sigma}$&$K^{*+}(892) \Lambda \phi$&$1$&$~g_{\Lambda^* \phi\Lambda}$&$g_{N^*K^* \Lambda}$\\
$K^{*0}(892) \Sigma^+ \rho^+$&$1$&$~\frac{1}{\sqrt{2}}g_{\Sigma^* \pi\Sigma}$&$\sqrt{\frac{2}{3}}g_{N^*K \Sigma}$&--&--&--&--\\
\hline
\end{tabular}
\end{table}}

%\clearpage
\section{Expressions for $N_{i,j}$, $\mathcal{D}_j$, $\mathcal{B}_{i,j}$ and $\mathcal{C}_{i,j}$}\label{ApB}
As mentioned in section~\ref{formalism}, the amplitudes for the different diagrams in Fig.~\ref{nstar}, as given by Eqs.~(\ref{ta2}), (\ref{tb2}) and (\ref{tc2}) are proportional to $N_{i,j}(~\vec {q}~)/\mathcal{D}_j(~\vec {q}~)$. The index $i$ indicates that  $N_{i,j}$ is the numerator resulting from the $q^0$ integration on  terms proportional to $(q^0)^i$. The index $j$ signifies that $N_{i,j}(~\vec {q}~)/\mathcal{D}_j(~\vec {q}~)$ is the result of the $q^0$ integration for the $j$th channel in the loop.  To facilitate writing the expressions of  $N_{i,j}$ and $\mathcal{D}_j$, we
 label the energies (masses) of the particles in the triangle loop with four-momentum $k-q$, $q$ and $P-k+q$ as, $E_{1j}$ ($m_{1j}$), $E_{2j}$ ($m_{2j}$) and $E_{Bj}$ ($m_{Bj}$), respectively, such that, in the center of mass frame:
\begin{align}
&E_{1j}=\sqrt{\left(\vec{k}-\vec{q}\right)^2+m_{1j}^2},\nonumber\\
&E_{2j}=\sqrt{\vec{q}^{\,\,2}+m_{2j}^2},\nonumber\\
&E_{Bj}=\sqrt{\left(-\vec{k}+\vec{q}\right)^2+m_{Bj}^2}.
\end{align}
Using the above definitions, we can write the numerators $N_{i,j}(\vec {q})$ as
\begin{align}
N_{0j}&=- E_{2j} \left(E_{Bj}+E_{1j}\right)\left(k^0\right)^2 + 2 \sqrt{s}E_{Bj} E_{2j} k^0+\left(E_{1j}+E_{2j}\right)\Biggl[E_{Bj}\left(E_{Bj}+E_{1j}-\sqrt{s}\right)\Biggr.\nonumber\\
&\times\left(E_{Bj}+E_{1j}+\sqrt{s}\right)+\Biggl.E_{2j}^2\left(E_{Bj}+E_{1j}\right)+E_{2j}\left(E_{Bj}+E_{1j}\right)\left(2E_{Bj}+E_{1j}\right)\Biggr],\label{n0}
\end{align}
\begin{align}
N_{1j}&= E_{2j}\Biggl\{E_{Bj}\Biggl[k^0\left(2E_{1j}^2+4E_{1j}E_{2j}+E_{2j}^2-s\right)-2E_{1j}\sqrt{s} \left(E_{1j}+E_{2j}\right) +2\sqrt{s}\left(k^0\right)^2 \!\!\Biggr.\Biggr.\nonumber\\
&\Biggl.\Biggl.-\! \left(k^0\right)^3\Biggr]\!+\!2E_{Bj}^2\left(E_{1j}+E_{2j}\right)k^0 \!+\!E_{Bj}^3K^0 -E_{1j}\left(\sqrt{s}-k^0\right)\!\left[\left(E_{1j}+E_{2j}\right)^2-\left(k^0\right)^2\right]\Biggr\},\label{n1}
\end{align}
\begin{align}
N_{2j}&=E_{2j}\Biggl\{\!E_{Bj}\!\Biggl[\left(k^0\right)^2\!\left(E_{1j}^2+4E_{1j}E_{2j}+E_{2j}^2-s\right)\!+\!E_{1j}\!\left(-E_{1j}^2E_{2j}+E_{1j}\left(s-E_{2j}^2\right)+sE_{2j} \right)\Biggr.\Biggr.\nonumber\\
&\Biggl.\!-2E_{1j}\sqrt{s}\left(\!E_{1j}+2E_{2j}\right)\!k^0\!+\!2\sqrt{s}\left(k^0\right)^3\!-\!\left(k^0\right)^4\!\Biggr]\!\!+\!E_{Bj}^2\!\left[\!\left(k^0\right)^2\!\left(E_{1j}+2E_{2j}\right)\!-\!E_{1j}\!\left(\!E_{1j}+2E_{2j}\right)^2\!\right]\nonumber\\
& +E_{Bj}^3\left[\left(k^0\right)^2-E_{1j}^2-E_{1j}E_{2j}\right]+E_{1j}\left(\sqrt{s}-k^0\right)^2\Biggr.\left[\left(E_{1j}+E_{2j}\right)^2-\left(k^0\right)^2\right]\Biggl\},\label{n2}
\end{align}
\begin{align}
N_{3j}&=E_{2j}\Biggl\{\!E_{Bj}\Biggl[\!\left(k^0\right)^3\!\left(E_{1j}^2+4E_{1j}E_{2j}\!+\!E_{2j}^2\!-\!s\right)\!+\!E_{1j}k^0\!\left(\!-2E_{1j}^2E_{2j}\!+\! E_{1j}\!\left(s-3E_{2j}^2\right)\!+\!2sE_{2j}\! \right)\Biggr.\Biggr.\nonumber\\
&\Biggl.+2E_{1j}^2E_{2j}\sqrt{s}\left(E_{1j}+E_{2j}\right)-2E_{1j}\sqrt{s}\left(E_{1j}+3E_{2j}\right)\left(k^0\right)^2+2\sqrt{s}\left(k^0\right)^4-\left(k^0\right)^5\Biggr]\Biggr.\nonumber\\
&\Biggl.+\!E_{Bj}^2\!\Biggl[\!-E_{1j}\!\left(\!E_{1j}^2\!+\!4E_{1j}E_{2j}\!+\!3E_{2j}^2\right)\!k^0\!+\!\left(\!E_{1j}\!+\!2E_{2j}\!\right)\!\left(\!k_0\!\right)^3\!+\!E_{1j}\sqrt{s}\left(\!E_{1j}\!+\!E_{2j}\right)^2\!-\!E_{1j}\sqrt{s}\left(k^0\right)^2\!\Biggr]\!\nonumber\\
&+\!E_{Bj}^3\left[\left(k^0\right)^3-E_{1j}\left(E_{1j}+2E_{2j}\right)k^0\right]\!-\!E_{1j}\left(\sqrt{s}-\left(k^0\right)\right)^3\left[\left(E_{1j}+E_{2j}\right)^2-\left(k^0\right)^2\right]\Biggl.\Biggr\},\!\label{n3}
\end{align}
\begin{align}
N_{4j}&=\!E_{2j}\!\Biggl\{\!2E_{Bj}^2\!\Biggl[\!-E_{1j}\!\left(k^0\right)^2\!\left(\!E_{1j}^2\!+\!4E_{1j}E_{2j}\!+\!3E_{2j}^2\!-\!s\!\right)\!+\!2E_{1j}\sqrt{s}k^0\!\left(\!E_{1j}\!+\!E_{2j}\! \right)^2\!+\!\left(E_{1j}\!+\!E_{2j}\right)\!\left(k^0\right)^4\!\Biggr.\Biggr.\nonumber\\
&+E_{1j}\left(E_{1j}+E_{2j}\! \right)^2\left(E_{1j}E_{2j}-s \right)\!-\!2E_{1j}\sqrt{s}\left(k^0\right)^3\!\Biggl.\Biggr]\!+\!E_{Bj}\!\Biggl[\left(k^0\right)^4\left(2E_{1j}^2+4E_{1j}E_{2j}+E_{2j}^2-s\right)\Biggr.\nonumber\\
&+\!E_{1j}\left(E_{1j}+E_{2j}\right)\!\left(E_{1j}^2\!\left[E_{2j}^2-s\right]-3sE_{1j}E_{2j}-sE_{2j}^2\right)\!+\!2E_{1j}^2\sqrt{s}\left(E_{1j}+2E_{2j}\right)^2k^0\!-\!E_{1j}\left(k^0\right)^2\nonumber\\
&\times\left(E_{1j}^3+4E_{1j}^2E_{2j}+6E_{1j}E_{2j}^2-2sE_{1j}-4sE_{2j}\right) -4E_{1j}\sqrt{s}\left(E_{1j}+2E_{2j}\right)\left(k^0\right)^3+2\sqrt{s}\left(k^0\right)^5\nonumber\\ 
&-\left(k^0\right)^6\Biggl.\Biggr]\!+\!E_{Bj}^3\Bigl[E_{1j}\left(E_{1j}^3+4E_{1j}^2E_{2j}+4E_{1j}E_{2j}^2+E_{2j}^3\right)-2E_{1j}\left(E_{1j}+2E_{2j}\right)\left(k^0\right)^2+\left(k^0\right)^4\Bigr]\nonumber\\
&+E_{Bj}^4E_{1j}\left[\left(E_{1j}+E_{2j}\right)^2-\left(k^0\right)^2\right]+E_{1j}\left(\sqrt{s}-k^0\right)^4\left[\left(E_{1j}+E_{2j}\right)^2-\left(k^0\right)^2\right]\Biggl.\Biggr\}.\label{n4}
\end{align}
The expression found for the denominator of Eq.~(\ref{nbyd}) is
\begin{align}
\mathcal{D}_j&=2E_{Bj}E_{1j}E_{2j}\left(\sqrt{s}-E_{Bj}-E_{1j}+i\epsilon\right)\left(\sqrt{s}+E_{Bj}+E_{1j}\right)\left(k^0-E_{1j}-E_{2j}+i\epsilon\right)\nonumber\\
&\times\left(k^0+E_{1j}+E_{2j}\right)\left(\sqrt{s}-k^0-E_{Bj}-E_{2j}+i\epsilon\right)\left(-\sqrt{s}+k^0-E_{Bj}-E_{2j}+i\epsilon\right),\label{deno}
\end{align}
where $i\epsilon$ is replaced by $i\Gamma/2$ for vector mesons with large widths, like $\rho$ and $K^*(892)$. We consider an average width for $\rho$ and $K^*(892)$ as 150 MeV and 50 MeV, respectively.

Further, we give the expressions of $\mathcal{B}_{i,j}$ needed to calculate Eq.~(\ref{tb2}),
\begin{align}
\mathcal{B}_{0,j}=&\vec \sigma \cdot \vec k \left\{-\left(m_{Bj}+m_{\Sigma^*}\right)\left(1+\frac{k^0}{E_{H^*}+m_{\Sigma^*}}\right)-\frac{2 \vec k \cdot \vec q+|\,\vec q\,|^2}{E_{H^*}+m_{\Sigma^*}}+\left(\frac{|~\vec k~|^2+|\,\vec q\,|^2}{\left(m_{vj}\right)^2}\right)\right.\nonumber\\
&\left.\times\left(\left(m_{Bj}+m_{\Sigma^*}\right)\left(1+\frac{k^0}{E_{H^*}+m_{\Sigma^*}}\right)+\frac{2 \vec k \cdot \vec q-|\,\vec q\,|^2}{E_{H^*}+m_{\Sigma^*}}\right)\right\}+\vec \sigma \cdot \vec q\Biggl\{k^0-m_{Bj}\nonumber\\
&\left.-m_{\Sigma^*}\!+\frac{|~\vec k~|^2}{E_{H^*}+m_{\Sigma^*}}+\left(\frac{|~\vec k~|^2+|\,\vec q\,|^2}{\left(m_{vj}\right)^2}\right)\!\left(-k^0-m_{Bj}-m_{\Sigma^*}-\frac{|~\vec k~|^2}{E_{H^*}+m_{\Sigma^*}}\right)\!\right\},\\
\mathcal{B}_{1,j}=&\vec \sigma \cdot \vec k \left\{-1+\frac{k^0-m_{Bj}-m_{\Sigma^*}}{E_{H^*}+m_{\Sigma^*}}+\left(\frac{|~\vec k~|^2+|\,\vec q\,|^2}{\left(m_{vj}\right)^2}\right)\left(1-\frac{k^0+m_{Bj}+m_{\Sigma^*}}{E_{H^*}+m_{\Sigma^*}}\right)\right\},
\end{align}
\begin{align}
\mathcal{B}_{2,j}=&\vec \sigma \cdot \vec k \left\{\frac{1}{E_{H^*}+m_{\Sigma^*}}-\frac{\left(m_{Bj}+m_{\Sigma^*}\right)}{\left(m_{vj}\right)^2}\left(1+\frac{k^0}{E_{H^*}+m_{\Sigma^*}}\right)+\frac{-2 \vec k \cdot \vec q+2|\,\vec q\,|^2+|~\vec k~|^2}{\left(E_{H^*}+m_{\Sigma^*}\right)\left(m_{vj}\right)^2}\right\}\nonumber\\
&+\frac{\vec \sigma \cdot \vec q}{\left(m_{vj}\right)^2}\left\{k^0+m_{Bj}+m_{\Sigma^*}\!+\frac{|~\vec k~|^2}{E_{H^*}+m_{\Sigma^*}}\right\},\\
\mathcal{B}_{3,j}=&\frac{\vec \sigma \cdot \vec k}{\left(m_{vj}\right)^2} \left\{-1+\frac{k^0+m_{Bj}+m_{\Sigma^*}}{E_{H^*}+m_{\Sigma^*}}\right\},\\
\mathcal{B}_{4,j}=&-\frac{\vec \sigma \cdot \vec k}{\left(E_{H^*}+m_{\Sigma^*}\right)\left(m_{vj}\right)^2}.
\end{align}

Finally, the terms $\mathcal{C}_{i,j}$, in Eq.~(\ref{tc2}), are
\begin{align}
\mathcal{C}_{0,j}=&\vec \sigma \cdot \vec k \left\{\vec k \cdot \vec q \left(\frac{\sqrt{s}+m_{Bj}+m_{\Sigma^*}}{E_{H^*}+m_{\Sigma^*}}\right)-\frac{|\,\vec q\,|^2 \left(\sqrt{s}+m_{\Sigma^*}\right)}{E_{H^*}+m_{\Sigma^*}}\right\}+\vec \sigma \cdot \vec q \left\{\vec k \cdot \vec q\right.\nonumber
\\
&\Biggl.-|~\vec k~|^2\left(\frac{\sqrt{s}+m_{Bj}+m_{\Sigma^*}}{E_{H^*}+m_{\Sigma^*}}\right)- k^0\left(m_{Bj}+m_{\Sigma^*}\right)\Biggr\},\\
\mathcal{C}_{1,j}=&\vec \sigma \cdot \vec k \left\{m_{Bj}+m_{\Sigma^*}+\frac{\vec k \cdot \vec q}{E_{H^*}+m_{\Sigma^*}}\right\}-\vec \sigma \cdot \vec q ~k^0,\\
\mathcal{C}_{2,j}=&\vec \sigma \cdot \vec k.
\end{align}


\begin{thebibliography}{}
%\cite{Tanabashi:2018oca}
\bibitem{pdg} 
  M.~Tanabashi {\it et al.} [ParticleDataGroup],
  %``Review of Particle Physics,''
  Phys.\ Rev.\ D {\bf 98}, no. 3, 030001 (2018).
%  doi:10.1103/PhysRevD.98.030001
  %%CITATION = doi:10.1103/PhysRevD.98.030001;%%
  %143 citations counted in INSPIRE as of 05 Sep 2018

%\cite{Khemchandani:2013nma}
\bibitem{Khemchandani:2013nma}
K.~Khemchandani, A.~Mart\'inez Torres, H.~Nagahiro and A.~Hosaka,
%``Role of vector and pseudoscalar mesons in understanding $1/2^- N^*$ and ? resonances,''
Phys.\ Rev.\ D \textbf{88}, no.11, 114016 (2013).
%doi:10.1103/PhysRevD.88.114016
%[arXiv:1307.8420 [nucl-th]].
%21 citations counted in INSPIRE as of 02 Apr 2020

 %\cite{Arndt:1995bj}
\bibitem{arndt} 
  R.~A.~Arndt, I.~I.~Strakovsky, R.~L.~Workman and M.~M.~Pavan,
  %``Updated analysis of pi N elastic scattering data to 2.1-GeV: The Baryon spectrum,''
  Phys.\ Rev.\ C {\bf 52}, 2120 (1995)
  [nucl-th/9505040].
  
  %\cite{Isgur:1978xj}
\bibitem{Isgur:1978xj}
N.~Isgur and G.~Karl,
%``P Wave Baryons in the Quark Model,''
Phys. Rev. D \textbf{18}, 4187 (1978)
doi:10.1103/PhysRevD.18.4187
  
  %\cite{Bijker:1994yr}
\bibitem{Bijker:1994yr}
R.~Bijker, F.~Iachello and A.~Leviatan,
%``Algebraic models of hadron structure. 1. Nonstrange baryons,''
Annals Phys. \textbf{236}, 69-116 (1994)
doi:10.1006/aphy.1994.1108
[arXiv:nucl-th/9402012 [nucl-th]].

  %\cite{Hosaka:1997kh}
\bibitem{Hosaka:1997kh}
A.~Hosaka, H.~Toki and M.~Takayama,
%``Baryon spectra in deformed oscillator quark model,''
Mod. Phys. Lett. A \textbf{13}, 1699-1708 (1998)
doi:10.1142/S0217732398001777
[arXiv:hep-ph/9711295 [hep-ph]].  

%\cite{Takayama:1999kc}
\bibitem{Takayama:1999kc}
M.~Takayama, H.~Toki and A.~Hosaka,
%``Systematics of the SU(3) baryon spectra and deformed oscillator quark model,''
Prog. Theor. Phys. \textbf{101}, 1271-1283 (1999)
doi:10.1143/PTP.101.1271
%9 citations counted in INSPIRE as of 28 Sep 2020

  \bibitem{osetramos}
  E.~Oset and A.~Ramos,
  %``Nonperturbative chiral approach to s wave anti-K N interactions,''
  Nucl.\ Phys.\ A {\bf 635} (1998) 99.

%\cite{Oller:2000fj}
\bibitem{Oller:2000fj} 
  J.~A.~Oller and U.~G.~Mei{\ss}ner,
  %``Chiral dynamics in the presence of bound states: Kaon nucleon interactions revisited,''
  Phys.\ Lett.\ B {\bf 500}, 263 (2001).
 % doi:10.1016/S0370-2693(01)00078-8
  %[hep-ph/0011146].
  %%CITATION = doi:10.1016/S0370-2693(01)00078-8;%%
  %669 citations counted in INSPIRE as of 17 Jul 2018

 %\cite{Jido:2003cb}
\bibitem{Jido:2003cb} 
  D.~Jido, J.~A.~Oller, E.~Oset, A.~Ramos and U.~G.~Mei{\ss}ner,
  %``Chiral dynamics of the two Lambda(1405) states,''
  Nucl.\ Phys.\ A {\bf 725}, 181 (2003).
 % doi:10.1016/S0375-9474(03)01598-7
  %[nucl-th/0303062].
  %%CITATION = doi:10.1016/S0375-9474(03)01598-7;%%
  %562 citations counted in INSPIRE as of 04 Sep 2018

%\cite{Hyodo:2011ur}
\bibitem{Hyodo:2011ur} 
  T.~Hyodo and D.~Jido,
  %``The nature of the Lambda(1405) resonance in chiral dynamics,''
  Prog.\ Part.\ Nucl.\ Phys.\  {\bf 67}, 55 (2012).
%  doi:10.1016/j.ppnp.2011.07.002
  %[arXiv:1104.4474 [nucl-th]].
  %%CITATION = doi:10.1016/j.ppnp.2011.07.002;%%
  %216 citations counted in INSPIRE as of 17 Jul 201

\bibitem{Mai:2014xna} 
  M.~Mai and U.~G.~Mei\ss ner,
  %``Constraints on the chiral unitary $\bar KN$ amplitude from $\pi\Sigma K^+$ photoproduction data,''
Eur.\ Phys.\ J.\ A {\bf 51}, no. 3, 30 (2015)
doi:10.1140/epja/i2015-15030-3
[arXiv:1411.7884 [hep-ph]].
%%CITATION = doi:10.1140/epja/i2015-15030-3;%%
  %87 citations counted in INSPIRE as of 20 May 2019



  %\cite{Kim:2017nxg}
\bibitem{Kim:2017nxg}
S.~Kim, S.~Nam, D.~Jido and H.~Kim,
%``Photoproduction of $\Lambda (1405)$ with the $N^*$ and $t$-channel Regge contributions,''
Phys.\ Rev.\ D \textbf{96}, no.1, 014003 (2017).
%doi:10.1103/PhysRevD.96.014003
%[arXiv:1702.08645 [hep-ph]].
%7 citations counted in INSPIRE as of 02 Apr 2020
%\cite{Oller:2000fj}

%\cite{Noumi:2017sdz}
\bibitem{Noumi:2017sdz}
H.~Noumi,
%``Strange and Charm Hadron Physics at J-PARC in Future,''
JPS Conf. Proc. \textbf{17}, 111003 (2017)
doi:10.7566/JPSCP.17.111003
%3 citations counted in INSPIRE as of 11 Sep 2020


\bibitem{Oller:2000fj} 
  J.~A.~Oller and U.~G.~Mei{\ss}ner,
  %``Chiral dynamics in the presence of bound states: Kaon nucleon interactions revisited,''
  Phys.\ Lett.\ B {\bf 500}, 263 (2001).
 % doi:10.1016/S0370-2693(01)00078-8

\bibitem{Guo}
Zhi-Hui Guo and J. A. Oller, Phys. Rev. C.~{\bf 87}, 035202 (2013).
  
  %\cite{Wu:2009tu}
\bibitem{Wu:2009tu} 
  J.~J.~Wu, S.~Dulat and B.~S.~Zou,
  %``Evidence for a new Sigma* resonance with J**P = 1/2- in the old data of K- p ---> Lambda pi+ pi- reaction,''
  Phys.\ Rev.\ D {\bf 80}, 017503 (2009).
%  doi:10.1103/PhysRevD.80.017503
  %[arXiv:0906.3950 [hep-ph]].
  %%CITATION = doi:10.1103/PhysRevD.80.017503;%%
  %30 citations counted in INSPIRE as of 17 Jul 2018
  
  %\cite{Wu:2009nw}
\bibitem{Wu:2009nw} 
  J.~J.~Wu, S.~Dulat and B.~S.~Zou,
  %``Further evidence for the Sigma* resonance with J**P = 1/2- around 1380-MeV,''
  Phys.\ Rev.\ C {\bf 81}, 045210 (2010).
%  doi:10.1103/PhysRevC.81.045210
  %[arXiv:0909.1380 [hep-ph]].
  %%CITATION = doi:10.1103/PhysRevC.81.045210;%%
  %32 citations counted in INSPIRE as of 17 Jul 2018
%\cite{Gao:2010hy}
\bibitem{Gao:2010hy} 
  P.~Gao, J.~J.~Wu and B.~S.~Zou,
  %``Possible $\Sigma({1\over2}^-)$ under the $\Sigma^*(1385)$ peak in $K\Sigma^*$ photoproduction,''
  Phys.\ Rev.\ C {\bf 81}, 055203 (2010).
%  doi:10.1103/PhysRevC.81.055203
  %[arXiv:1001.0805 [nucl-th]].
  
  %\cite{Xie:2014zga}
\bibitem{Xie:2014zga} 
  J.~J.~Xie, J.~J.~Wu and B.~S.~Zou,
  %``Role of the possible $\Sigma^*(\frac{1}{2}^-)$ state in the $\Lambda p \to \Lambda p \pi^0$ reaction,''
  Phys.\ Rev.\ C {\bf 90}, no. 5, 055204 (2014).
%  doi:10.1103/PhysRevC.90.055204
  %[arXiv:1407.7984 [nucl-th]].
  %%CITATION = doi:10.1103/PhysRevC.90.055204;%%
  %7 citations counted in INSPIRE as of 17 Jul 2018

  
  \bibitem{Xie:2017xwx} 
  J.~J.~Xie and L.~S.~Geng,
  %``$\Sigma^*_{1/2^-}(1380)$ in the $\Lambda^+_c \to \eta \pi^+ \Lambda$ decay,''
  Phys.\ Rev.\ D {\bf 95}, no. 7, 074024 (2017).
%  doi:10.1103/PhysRevD.95.074024
  %[arXiv:1703.09502 [hep-ph]].
  %%CITATION = doi:10.1103/PhysRevD.95.074024;%%
  %2 citations counted in INSPIRE as of 17 Jul 2018

  
  \bibitem{Khemchandani:2012ur} 
  K.~P.~Khemchandani, A.~Martinez Torres, H.~Nagahiro and A.~Hosaka,
  %``Negative parity $\Lambda$ and $\Sigma$ resonances coupled to pseudoscalar and vector mesons,''
  Phys.\ Rev.\ D {\bf 85}, 114020 (2012).
%  doi:10.1103/PhysRevD.85.114020
  %[arXiv:1203.6711 [nucl-th]].
  %%CITATION = doi:10.1103/PhysRevD.85.114020;%%
  %25 citations counted in INSPIRE as of 17 Jul 2018
  
  %\cite{Roca:2013cca}
\bibitem{Roca:2013cca} 
  L.~Roca and E.~Oset,
  %``Isospin 0 and 1 resonances from $\pi \Sigma$ photoproduction data,''
  Phys.\ Rev.\ C {\bf 88}, 055206 (2013).
  %%CITATION = doi:10.1103/PhysRevC.88.055206;%%
  %45 citations counted in INSPIRE as of 19 Oct 2018

%\cite{Khemchandani:2018amu}
\bibitem{Khemchandani:2018amu}
K.~Khemchandani, A.~Mart\'inez Torres and J.~Oller,
%``Hyperon resonances coupled to pseudoscalar- and vector-baryon channels,''
Phys.\ Rev.\ C \textbf{100}, no.1, 015208 (2019).
%doi:10.1103/PhysRevC.100.015208
%[arXiv:1810.09990 [hep-ph]].
%4 citations counted in INSPIRE as of 02 Apr 2020


\bibitem{Mai:2014xna} 
  M.~Mai and U.~G.~Mei§ner,
  %``Constraints on the chiral unitary $\bar KN$ amplitude from $\pi\Sigma K^+$ photoproduction data,''
Eur.\ Phys.\ J.\ A {\bf 51}, no. 3, 30 (2015).
%doi:10.1140/epja/i2015-15030-3

\bibitem{Lu:2013nza} 
  H.~Y.~Lu {\it et al.} [CLAS Collaboration],
  %``First Observation of the $\Lambda(1405)$ Line Shape in Electroproduction,''
  Phys.\ Rev.\ C {\bf 88}, 045202 (2013).
  
  \bibitem{Bando:1984ej} 
  M.~Bando, T.~Kugo, S.~Uehara, K.~Yamawaki and T.~Yanagida,
  %``Is rho Meson a Dynamical Gauge Boson of Hidden Local Symmetry?,''
  Phys.\ Rev.\ Lett.\  {\bf 54}, 1215 (1985).
  
\bibitem{Bando:1987br} 
  M.~Bando, T.~Kugo and K.~Yamawaki,
  %``Nonlinear Realization and Hidden Local Symmetries,''
  Phys.\ Rept.\  {\bf 164}, 217 (1988).

%\cite{Oller:2018zts}
\bibitem{Oller:2018zts}
J.~Oller and D.~Entem,
%``The exact discontinuity of a partial wave along the left-hand cut and the exact $N/D$ method in non-relativistic scattering,''
Annals Phys.\  \textbf{411}, 167965 (2019).
%doi:10.1016/j.aop.2019.167965
%[arXiv:1810.12242 [hep-ph]].
%4 citations counted in INSPIRE as of 08 Apr 2020
%Copy to ClipboardDownload

%\cite{Moriya:2013hwg}
\bibitem{Moriya:2013hwg}
K.~Moriya \textit{et al.} [CLAS],
%``Differential Photoproduction Cross Sections of the $\Sigma^0(1385)$, $\Lambda(1405)$, and $\Lambda(1520)$,''
Phys. Rev. C \textbf{88}, 045201 (2013)
doi:10.1103/PhysRevC.88.045201
[arXiv:1305.6776 [nucl-ex]].
%84 citations counted in INSPIRE as of 10 Sep 2020


%\cite{Scheluchin:2020mhn}
\bibitem{Scheluchin:2020mhn}
G.~Scheluchin, S.~Alef, P.~Bauer, R.~Beck, A.~Braghieri, P.~Cole, R.~Di Salvo, D.~Elsner, A.~Fantini, O.~Freyermuth, F.~Ghio, A.~Gridnev, D.~Hammann, J.~Hannappel, T.~Jude, K.~Kohl, N.~Kozlenko, A.~Lapik, P.~Levi Sandri, V.~Lisin, G.~Mandaglio, R.~Messi, D.~Moricciani, V.~Nedorezov, D.~Novinsky, P.~Pedroni, A.~Polonski, B.~E.~Reitz, M.~Romaniuk, H.~Schmieden, V.~Sumachev, V.~Tarakanov and C.~Tillmanns,
%``$K^+\Lambda$(1405) photoproduction at the BGO-OD experiment,''
[arXiv:2007.08898 [nucl-ex]].
%0 citations counted in INSPIRE as of 10 Sep 2020
\end{thebibliography}
\end{document}